\documentclass[journal=jctcce,manuscript=article]{achemso}

\usepackage{chemformula} 
\usepackage[T1]{fontenc} 
\usepackage{amsfonts,latexsym,eucal,color,graphicx,epsfig,amssymb,amsmath,algorithm,algpseudocode,appendix}
\usepackage[caption=false]{subfig}
\algrenewcommand\algorithmicrequire{\textbf{Input}}
\algrenewcommand\algorithmicensure{\textbf{Output}}

\newcommand{\vqe}{\mathrm{VQE}}
\newcommand{\qpe}{\mathrm{QPE}}

\author{Nick S. Blunt}
\altaffiliation{These authors contributed equally to this work.}
\author{Joan Camps}
\altaffiliation{These authors contributed equally to this work.}
\author{Ophelia Crawford}
\altaffiliation{These authors contributed equally to this work.}
\author{R\'obert Izs\'ak}
\altaffiliation{These authors contributed equally to this work.}
\author{Sebastian Leontica}
\altaffiliation{These authors contributed equally to this work.}
\author{Arjun Mirani}
\altaffiliation{These authors contributed equally to this work.}
\author{Alexandra E. Moylett}
\altaffiliation{These authors contributed equally to this work.}
\author{Sam A. Scivier}
\altaffiliation{These authors contributed equally to this work.}
\author{Christoph Sünderhauf}
\altaffiliation{These authors contributed equally to this work.}
\affiliation{Riverlane, St Andrews House, 59 St Andrews Street, Cambridge, CB2 3BZ,  UK}
\author{Patrick Schopf}
\affiliation{Astex Pharmaceuticals, 436 Cambridge Science Park, Cambridge, CB4 0QA, UK}
\author{Jacob M. Taylor}
\author{Nicole Holzmann}
\affiliation{Riverlane, St Andrews House, 59 St Andrews Street, Cambridge, CB2 3BZ,  UK}
\alsoaffiliation{Astex Pharmaceuticals, 436 Cambridge Science Park, Cambridge, CB4 0QA, UK}
\email{nicole.holzmann@riverlane.com}

\title[Pharma Costing]{A perspective on the current state-of-the-art of quantum computing for drug discovery applications}


\begin{document}


\begin{abstract}
Computational chemistry is an essential tool in the pharmaceutical industry. Quantum computing is a fast evolving technology that promises to completely shift the computational capabilities in many areas of chemical research by bringing into reach currently impossible calculations.\\
This perspective illustrates the near-future applicability of quantum computation of molecules to pharmaceutical problems. We briefly summarize and compare the scaling properties of state-of-the-art quantum algorithms, and provide novel estimates of the quantum computational cost of simulating progressively larger embedding regions of a pharmaceutically relevant covalent protein-drug complex involving the drug Ibrutinib. Carrying out these calculations requires an error-corrected quantum architecture, that we describe. Our estimates showcase that recent developments on quantum phase estimation algorithms have dramatically reduced the quantum resources needed to run fully quantum calculations in active spaces of around 50 orbitals and electrons, from estimated over 1000 years using the Trotterisation approach to just a few days with sparse qubitisation, painting a picture of fast and exciting progress in this nascent field.
\end{abstract}


\section{Introduction}
The drug design process is a complex procedure in which computers and wet-lab methods are used together in pursuing new pharmaceuticals. Although most methods of computer-aided drug design (CADD) rely on statistical fitting methods or on classical mechanics \cite{yu2017computer}, it has been argued that more accurate quantum mechanical methods have an important contribution to make to several aspects of CADD \cite{tautermann2020current, lam2020applications, heifetz_quantum_2020}. Unfortunately, finding exact or nearly exact solutions for chemically relevant systems becomes intractable for more than $\sim$30 electrons\cite{hermann_deep-neural-network_2020}. Although efficient and accurate approximations exist for much larger systems than that, it remains desirable to find methods that can deliver the exact quantum mechanical solution in a cost-efficient way.

Being quantum systems themselves, quantum computers are naturally suited to simulating quantum mechanical problems without running out of memory exponentially fast.  Many aspects of chemical research are expected to benefit from accurate quantum methods, including in the pharmaceutical industry \cite{BCG_report, McKinsey_report}. Carrying out such industrially disruptive quantum simulations requires very high fidelity quantum computers.

Quantum computers have seen a significant number of experimental developments over the last several years. Recent trapped ion quantum devices have average two-qubit gate fidelities of up to 99.8\% \cite{quantinuum_fidelity}. Developments have also been seen in superconducting technologies, most famously shown in the 50-60 qubit devices from Google and USTC which claim to show quantum computational advantage, the point where a quantum computer is believed to have solved a classically intractable problem, albeit not a problem with applications in quantum chemistry \cite{arute_quantum_2019, wu_strong_2021}. Superconducting devices have also been developed with over 120 qubits, in particular IBM's Eagle processor \cite{collins2021ibm}. These devices are now at a point where it is possible to run Noisy Intermediate-Scale Quantum algorithms such as the Variational Quantum Eigensolver \cite{peruzzo_variational_2014}, with the largest experimental efforts to date simulating the binding energy of hydrogen chains with up to 12 atoms \cite{hempel_quantum_2018, chen_demonstration_2020, google_ai_quantum_and_collaborators_hartree-fock_2020}. Along with early applications, experimental groups have started showing initial implementations of quantum error correction, a fundamental step in scaling up quantum computing where multiple physical qubits are used to protect a smaller number of logical qubits from errors \cite{roffe_quantum_2019}, thus increasing the effective fidelity. These experiments have been shown to suppress errors while keeping a single logical qubit alive and applying some simple logical gates in a number of platforms, including superconducting devices \cite{chen_exponential_2021}, trapped ions \cite{nguyen_demonstration_2021, egan_fault-tolerant_2021, pino_demonstration_2021, postler2021demonstration}, and nuclear-spin qubits in diamond \cite{abobeih_fault-tolerant_2021}. These results show the significant progress that has been made in quantum hardware, as well as laying the groundwork to reaching large-scale fault-tolerant quantum computation.

This perspective focuses on the disruption enabled by the large size complete active space configuration interaction (CASCI) calculations admitted by near-future quantum computers. We discuss the steps involved in running a pharmaceutical application on a quantum computer, from mapping the chemical problem onto quantum memory, selecting a quantum algorithm, and specifying an error-corrected quantum architecture to solve it.  We illustrate these steps with an example system: the drug Ibrutinib bound covalently to Bruton’s tyrosine kinase. We estimate the quantum computational resources needed to fully quantum simulate progressively larger clusters of the binding pocket and the Ibrutinib inhibitor. Our estimates exhibit that quantum algorithmic developments over the past five years have dramatically reduced the quantum resources needed to run fully quantum calculations in active spaces of around 50 orbitals, which could be performed on sufficiently large error-corrected quantum computers with a runtime of just a few days.

This perspective is organised as follows. Sec.~\ref{sec:intro} discusses the mapping of the electronic structure problem onto a quantum computer. Sec.~\ref{sec:alg_scaling} discusses and compares the scaling of the two salient quantum algorithms for finding the ground-state energy of an electronic Hamiltonian -- VQE and QPE. We conclude that QPE scales more favourably, and the rest of the paper focuses on this algorithm. Sec.~\ref{sec:errorcorrection} discusses the aspects of error correction needed for estimating the quantum resources needed to run QPE. The main ingredient of the QPE algorithm is an efficiently-implemented unitary operator that is related to the Hamiltonian. Sec.~\ref{sec:trot_vs_qub} discusses two methods to construct such unitary operators: Trotterisation and qubitisation. Sec.~\ref{sec:chemical_system} discusses the pharmaceutical system of focus, the computational methods, and the active spaces used. Sec.~\ref{sec:results} contains the results of our resource estimations. We find that qubitisation gives much more favourable runtimes than Trotterisation. We conclude in Sec.~\ref{sec:outlook}.

\section{Chemistry on a quantum computer}
\label{sec:intro}

\subsection{Chemistry and the electronic structure problem}
\label{sec:chemistry_intro}

The question of how particles interact had already led the ancient Greek and Roman atomists to talk about ``hooked atoms'' that could intertwine and hold matter together. After atomism was revived two millennia later, in his Opticks, Newton preferred to hypothesise an attractive force, as yet unknown, that holds atoms together. Accumulating knowledge on electricity and electrochemistry in the 19th century favored explanations featuring electrostatic interactions in this role. The period also saw the rise of the theory of chemical valency that sought to determine the number of partners an atom might have in a compound and eventually led to the characterization of the combining forces as chemical bonds \cite{partington1989short}. After the discovery of the electron and the refinement of atom models that culminated in Bohr's model in 1913, G.N. Lewis put forward his own interpretation of the (covalent) chemical bond as electron pairs shared between atomic nuclei \cite{lewis1916atom} and to give a physical picture of that ``hook and eye'', as he put it \cite{lewis1923valence}. But the real breakthrough promising quantitative predictions came with the early application of quantum mechanics to simple chemical systems in the late 1920s, including the works of Burrau on H$_2^+$ \cite{burrau1927berechnung}, Heitler and London \cite{heitler1927wechselwirkung}, and later Pauling \cite{pauling1931nature} anticipating valence bond theory, with Hund \cite{hund1927deutung}, Mulliken \cite{mulliken1928assignment} and Lennard-Jones \cite{lennard-jones1929electronic} laying the foundations for molecular orbital theory. However, starting from the first principles of quantum mechanics and special relativity leads to equations that are insoluble in all but the simplest of cases, as Dirac lamented in 1929, concluding that more efficient approximate solutions are necessary \cite{dirac1929quantum}. Current wavefunction based methods of quantum chemistry rely on a series of approximations that lead to a computable first guess and then, whenever possible, various other methods are applied to account for the approximations made or, as computational chemists say, to correct for the various ``effects'' neglected. To begin with,  typically a single molecule in a vacuum is considered, without taking relativity into account and considering static solutions only. To facilitate a quantum mechanical treatment, one must be able to represent the interactions among the electrons and nuclei of molecules as a linear Hermitian operator, the Hamiltonian,  the eigensolutions of which represent the possible states of the system and the total energies associated with them. To simplify the problem further, the Born-Oppenheimer approximation \cite{born1927quantentheorie} posits that electronic and nuclear degrees of freedom can be separated, after which most methods focus on tackling the electronic problem. The resulting time-independent Schr\"{o}dinger equation reads
\begin{equation}
\hat{H}\Psi_k = E_k\Psi_k, \quad
\hat{H} = \hat{T}_e + \hat{V}_{ee} + \hat{V}_{ne} + \hat{V}_{nn},
\end{equation}
where $\Psi_k$ and $E_k$ are the electronic wavefunction and energy of the $k$th state, and $\hat{H}$ is the electronic Hamiltonian consisting of a kinetic energy term of the electrons ($\hat{T}_e$) and the potential energy terms of the electron-electron ($\hat{V}_{ee}$), nuclear-electron ($\hat{V}_{ne}$) and nuclear-nuclear ($\hat{V}_{nn}$) interactions.

At this stage, the computational problem is still intractable. Further progress was made by assuming that electronic coordinates can also be separated and the total wavefunction has the form of a Slater determinant \cite{slater1929theory}
\begin{equation}
\Phi = \hat{\mathcal{A}} \prod_{i=1}^{N}\varphi_i(i),
\end{equation}
where the antisymmetrizer $\hat{\mathcal{A}}$ permutes the particle labels and sums over terms with the appropriate sign and norm factor. Thus, the exact wavefunction describing $N$ electrons is approximated as a determinant $\Phi$ constructed from functions describing a single electron, the molecular (spin-)orbitals $\varphi_i$. Calculating the expectation value of $\hat{H}$ using $\Phi$ and minimizing it with respect to the orbitals yields the Hartree-Fock (HF) equations \cite{hartree1928wave,slater1930note,fock1930naeherungen}. To make the parametrization of this problem easier, the molecular orbitals themselves are expanded as a linear combination of known atomic orbitals. In molecular calculations, a convenient choice for the latter is Gaussian functions and, using these, the Hartree-Fock equations are reduced to a set of algebraic equations for the expansion coefficients of molecular orbitals \cite{roothaan1951new}. While the algebraic Hartree-Fock problem is soluble for molecules containing several hundred atoms, being an effective one-electron theory, it does not account for correlation effects between multiple electrons \cite{knowles2000ab}. However, once the molecular orbitals are obtained in a given atomic orbital basis, a linear combination of all possible Slater determinants will yield the exact solution in that basis. Unfortunately, this full configuration interaction (FCI) solution scales exponentially with the number of electrons and orbitals in the system. The classical solution to the problem is to define less expensive ans\"{a}tze for the wavefunction that only scale polynomially \cite{knowles2000ab}, e.g., the coupled cluster singles and doubles (CCSD) ansatz. When correlation effects are weak, i.e., when HF is a good starting guess, this approach has been extremely successful in many areas of chemistry. For strongly correlated systems, the most straightforward alternative to HF is obtaining the FCI solution within a complete active space (CAS) \cite{roos1980complete} rather than for the entire orbital space. We will refer to the configuration interaction solution within this active space (for our purposes, without orbital optimization) as CASCI. Unfortunately, this still leaves many important problems outside the reach of quantum chemical methods on the classical computer. For strongly correlated systems, the main difficulty lies in the size of the active spaces required for correctly describing some systems, an area where quantum computers may make a breakthrough \cite{reiher_elucidating_2017}. For weakly correlated systems, high quality results delivered by quantum computers may still yield significant improvements over popular density functional theory (DFT) approaches \cite{genin2022estimating} or even efficient wavefunction based approaches on the classical computer.

As it was pointed out even in the case of an archetypal strongly correlated complex \cite{reiher_elucidating_2017}, outperforming popular density functional methods is an important practical criterion. In contrast, usual definitions of quantum advantage involve formal criteria such as exponential speed-up, the relevance of which to chemistry has been recently questioned \cite{lee2022there}. Here, a less ambitious working definition is used: quantum benefit is reached if quantum computers can outperform classical computers in some industrially relevant process. An important part of that is the ability to give better results than DFT at a reasonable cost. Recent work on photochemical processes with simulated quantum computing uses precisely such criteria \cite{genin2022estimating}. Unfortunately, the assessment of potential quantum benefit for pharma remains a difficult task, not in the least because predictive application of quantum mechanics in the drug discovery process is a relatively new trend even using classical computers \cite{lam2020applications}. Some areas where quantum effects are known to be important, such as the description of weak hydrogen bonds \cite{anighoro2020underappreciated}, also stand out as the most likely candidates for quantum benefit. In this perspective, our aim is to provide quantum resource estimates for a protein-drug system in which such interactions play an important role.

\subsection{Quantum computation}
\label{sec:quantum_computation}

Quantum computers are computational devices that use the laws of quantum mechanics to perform calculations. The theory of quantum computing was first developed in the early 1980's by pioneers including Paul Benioff, Richard Feynman, David Deutsch and Peter Shor\cite{shor_early_quantum}. The motivation for quantum computing comes from the potential to perform calculations efficiently, which can only be performed inefficiently on a digital computer. Here, efficient means that the runtime is polynomial in the size of the problem.

This initial work led in 1994 to the development of Shor's algorithm\cite{shor1994}, which allows the prime factorisation of an integer to be performed in polynomial time, compared to the super-polynomial time required by classical algorithms. In 1996, Lov Grover developed an algorithm to search an unsorted database of size $N$ in $\mathcal{O}(\sqrt{N}$) time, compared to the $\mathcal{O}(N)$ runtime required by classical algorithms\cite{grover1996}. These discoveries demonstrate the potential for improved performance of certain quantum algorithms over classical ones.

The key to efficiently studying chemistry on a quantum computer came in the late 1990's. Alexei Kitaev, building on the work of Shor, introduced quantum phase estimation (QPE) in 1995 to study the Abelian stabiliser problem\cite{kitaev_quantum_1995}. In 1998, Cleve \emph{et al.} extended this QPE approach to estimate the phase of an arbitrary unitary operator\cite{cleve1998}; the form of QPE introduced here is identical to that often considered today. The QPE method can be applied to find the eigenvalues of a chemical Hamiltonian to a given accuracy with a runtime that scales polynomially with system size. For this reason, we believe that quantum computers can perform accurate chemical calculations beyond the reach of classical devices.

A quantum computer consists of a register of qubits, or quantum bits. Each of these qubits can be in a state $\vert 0 \rangle$ or $\vert 1 \rangle$. However, following the laws of quantum mechanics, the state can also be an arbitrary superposition of the two
\begin{equation}
\vert \Psi \rangle = \alpha \vert 0 \rangle + \beta \vert 1 \rangle,
\end{equation}
in addition to possible entanglement between the qubits. Time evolution in quantum mechanics is unitary, and as such the gates performed on the qubits are unitary operations too. In particular, a quantum computer is built to perform a small set of basis unitary operations at the physical level. These operations are designed to be universal; that is, any unitary operator on any number of qubits can be built from these basis gates. This can be achieved using gates that only act on one or two qubits at a time, a fact that is crucial for physical realisations of quantum computers; it is not realistic to perform physical operations that entangle large numbers of qubits simultaneously with high fidelity. Instead, these operations can be built from much simpler physical operations. Finally, state preparation and measurement are important components of quantum computation; qubits are each prepared in state $\vert 0 \rangle$ at the start of a computation, and measurement causes wave function collapse according to the Born rule.

One set of universal gates, which will be important for later discussion of quantum error correction, consists of the Hadamard gate ($H$), phase gates $S$ and $T$, defined in matrix form by
\begin{equation}
H = \frac{1}{\sqrt{2}}\begin{pmatrix}
1 & 1 \\
1 & -1 \\
\end{pmatrix}, \,\,\,\,\,\,\,\,
S = \begin{pmatrix}
1 & 0 \\
0 & i \\
\end{pmatrix}, \,\,\,\,\,\,\,\,
T = \begin{pmatrix}
1 & 0 \\
0 & e^{i\pi/4} \\
\end{pmatrix}
\label{eq:h_s_t_gates}
\end{equation}
and the CNOT gate, defined by
\begin{equation}
\mathrm{CNOT} = \begin{pmatrix}
1 & 0 & 0 & 0 \\
0 & 1 & 0 & 0 \\
0 & 0 & 0 & 1 \\
0 & 0 & 1 & 0 \\
\end{pmatrix}
\label{eq:cnot_gate}
\end{equation}
which flips the state of a second qubit, conditional on the first being in state $\vert 1 \rangle$. Programs built from these gates are often represented by circuit diagrams. An example of the QPE circuit diagram is given later in Sec.~\ref{sec:QPE}. For a thorough introduction to quantum computing, including circuit diagram notation, we refer the reader to Ref.~\cite{nielsen_quantum_2010}.

\subsection{The qubit Hamiltonian}
\label{sec:qubit_hamiltonian}

As discussed above, the great promise of quantum computers for chemistry is that they can find eigenvalues of a Hamiltonian with polynomial scaling. This would render a great number of strongly correlated chemical problems amenable to exact quantum mechanical treatment with potential benefits in many branches of the chemical industry \cite{reiher_elucidating_2017}. To realize this promise, the Hamiltonian encoding the interactions in the chemical system needs to be represented in a way that the quantum computer will be able to interpret. One possibility for this is the second quantised representation
\begin{equation}
\hat{H} = h_0 + \sum_{p, q}h^q_p a_p^{\dagger} a_q + \frac{1}{2} \sum_{p, q, r, s}h^{sq}_{pr} a_{p}^{\dagger} a_{r}^{\dagger} a_{q} a_{s},
\end{equation}
where the fermionic annihilation ($a_q$) and creation ($a_p^{\dagger}$) operators are summed over the molecular spin-orbital labels $p,q,\ldots$ within the active space. It is important to emphasize that such active spaces are often chosen to reduce the cost of the calculation and involve the projection of the full Hamiltonian to the CAS space \cite{dyall1995choice}. This causes screening terms to appear in the matrix elements: the constant term $h_0$ contains the nuclear-nuclear interaction and any screening terms, the one-body term $h^q_p$ includes the kinetic and nuclear-electron attraction as well as any screening terms, and the two-body term consists of the interelectronic repulsion term. Once the Hartree-Fock solution or some other set of molecular orbitals is available to define $p,q,\ldots$, it is possible to generate $h_0$, $h^q_p$ and $h^{qs}_{pr}$. The result of a quantum computation using such a CAS Hamiltonian corresponds to a CASCI calculation on the classical computer that becomes identical with the exact (FCI) solution in the limit that the active space includes the entire orbital space in a given basis.

In the next step, the fermionic operators need to be mapped to qubit operators, whose action on the qubits can be directly calculated. The Jordan-Wigner transformation \cite{jordan1928ueber} achieves this using Pauli spin-matrices and requiring that the new representation satisfy the anticommutation rules of fermion operators. The resulting transformation for creation operators reads
\begin{equation}
a_p^{\dagger} \quad\to\quad  \frac{1}{2}(X_p - iY_p)\bigotimes_{q<p} Z_{q},
\end{equation}
and for annihilation operators, it is
\begin{equation}
a_p \quad\to\quad  \frac{1}{2}(X_p + iY_p)\bigotimes_{q<p} Z_{q},
\end{equation}
where $X_p$, $Y_p$ and $Z_p$ are Pauli spin operators acting on the $p$th qubit.  It should be noted that there are alternatives to the Jordan-Wigner transformation, and it has recently been argued that for larger chemical problems, the one proposed by Bravyi and Kitaev \cite{Bravyi2002,seeley_bravyi-kitaev_2012} will be more advantageous \cite{tranter_comparison_2018}. Whichever method one chooses, the result is a qubit Hamiltonian, i.e., a linear combination of Pauli-strings that represent the chemical system for the quantum computer and serve as a starting point for quantum algorithms. We write this as
\begin{equation}
    {\hat{H}} = \sum_{i=1}^{L} w_{i} P_{i}, \label{eq:qubham}
\end{equation}
where each $P_{i}$ is a Pauli operator and $w_{i}$ its corresponding (real) coefficient.

There are two main classes of algorithm for performing computational chemistry calculations on quantum computers -- the variational quantum eigensolver (VQE) and quantum phase estimation (QPE). The focus of this paper is using the latter to estimate the quantum computational resources required to perform pharmaceutically-relevant chemistry calculations.

\subsection{Algorithms}
\label{sec:algorithms}
In this section, we outline the VQE and QPE algorithms.

\subsubsection{The variational quantum eigensolver}
\begin{figure}
    \centering
    \includegraphics[width=0.9\textwidth]{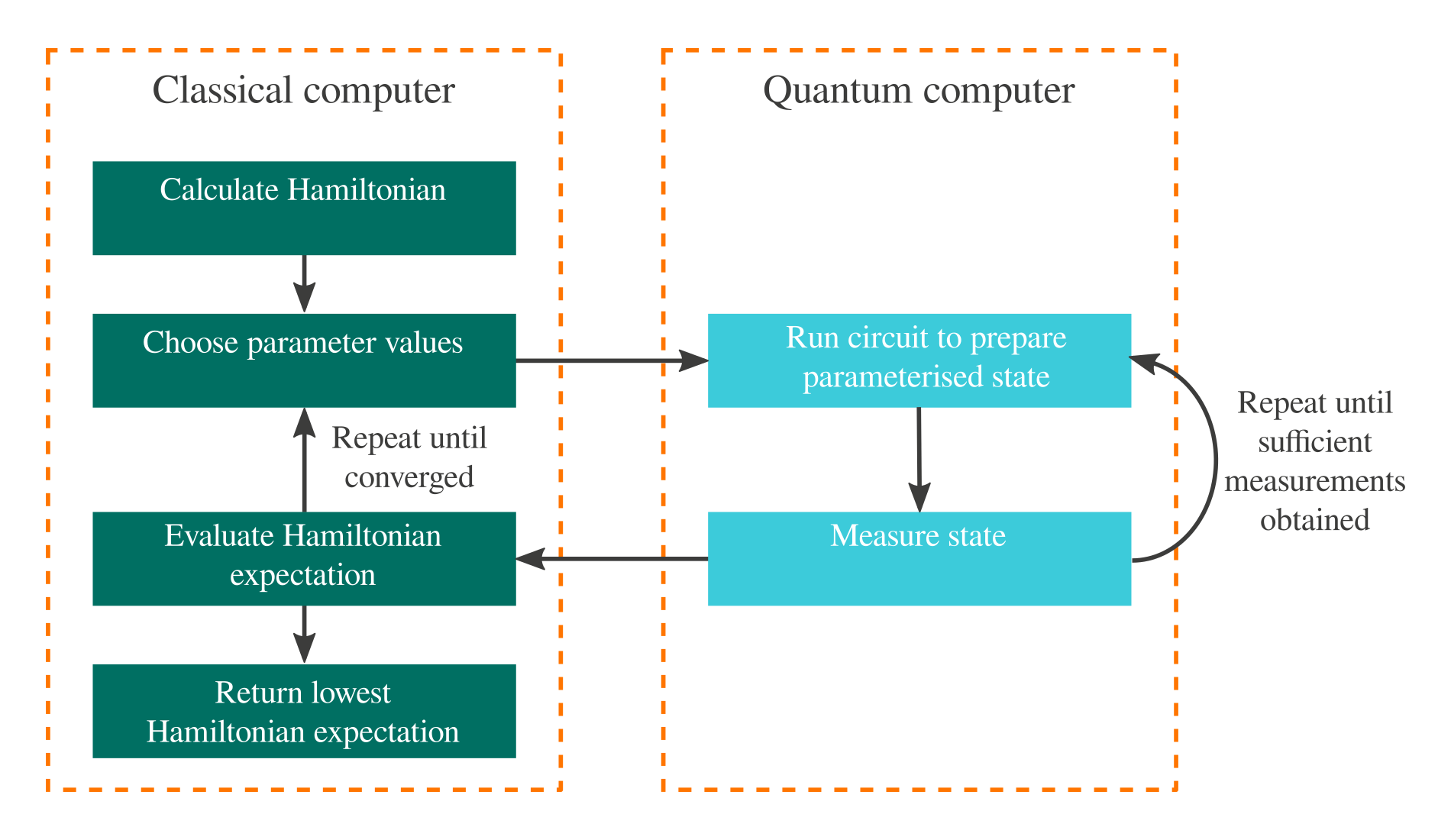}
    \caption{Outline of the VQE algorithm, indicating which parts occur on the quantum computer and which parts on the classical.}
    \label{fig:vqe}
\end{figure}
VQE~\cite{peruzzo_variational_2014} is a hybrid algorithm, making use of both classical and quantum computational resources. A classical optimiser explores some set of quantum states, seeking that with the smallest Hamiltonian expectation value. By the variational principle, any such expectation value is necessarily greater than or equal to the ground-state energy. It is therefore hoped that the smallest expectation value will be close to the ground-state energy. Excited-state energies can also be sought through extensions (e.g. ~\cite{higgott_variational_2019, mcclean_hybrid_2017, motta2020determining,zhang2020variational,ryabinkin_constrained_2018}).

The set of states explored is known as an ansatz. These states are prepared through some parameterised quantum circuit. Having chosen some initial parameter values, the ansatz circuit is run to prepare a particular quantum state and measurements of the state made. Typically the ansatz circuit must be applied many times to obtain sufficient information to estimate the expectation of the Hamiltonian on the quantum state to some desired level of accuracy. Based on this expectation, the parameter values are updated by the classical optimiser and the expectation estimation process repeated until some convergence criteria are satisfied. The VQE algorithm is illustrated in Fig.~\ref{fig:vqe} and we provide further details of the different aspects of the algorithm in section~\ref{sec:vqeresources}.

\subsubsection{Quantum phase estimation}
\label{sec:QPE}
\begin{figure}
    \centering
    \includegraphics[width=0.9\textwidth]{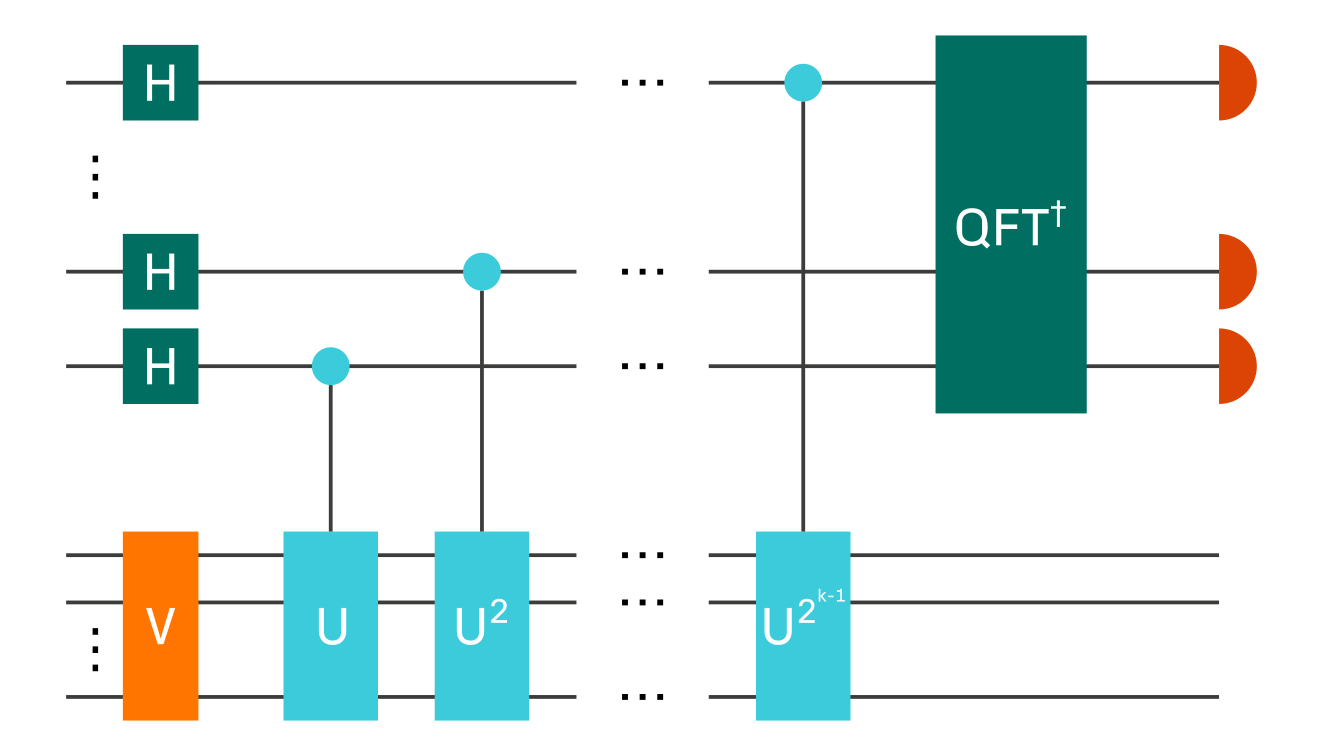}
    \caption{Outline of the circuit used to perform QPE, adapted from~\cite{nielsen2002quantum}.}
    \label{fig:qpe}
\end{figure}
Quantum phase estimation is another algorithm for calculating energies of chemical systems~\cite{kitaev_quantum_1995, kitaev2002classical}. It requires deeper circuits than VQE, but these must be performed typically only a handful of times. Furthermore, QPE does not require an ansatz; it instead calculates eigenvalues of the Hamiltonian directly, up to some level of precision.

QPE makes use of the quantum Fourier transform to estimate the eigenphases of a unitary operator, $U$. An eigenphase, $\varphi_i$, satisfies
\begin{equation}
    U \vert \Psi_{i} \rangle  = e^{i \varphi_{i}} \vert \Psi_{i} \rangle,
\end{equation}
where $\vert \Psi_{i} \rangle$ is the corresponding eigenstate. In order to perform computational chemistry calculations, the unitary must be constructed from the Hamiltonian; one choice is $U=e^{-i{\hat{H}}t}$. The operators $U$ and ${\hat{H}}$ share eigenstates and their eigenvalues are related through $E_{i}t = -\varphi_{i}$. An outline of the circuit used to perform QPE is shown in Fig.~\ref{fig:qpe}. There are two sets of qubits; the state register (bottom) used to prepare the eigenstate by the end of the calculation and the data register (top) used to read bits corresponding to $\frac{\varphi_{i}}{2\pi}$ in binary fraction representation. The precision of the estimate of the energy is thus limited by the number of qubits in the data register.

Initially, the state register is prepared to contain a state which is hoped to have significant overlap with the true ground state -- a common choice when performing chemistry calculations is the Hartree-Fock state. A sequence of controlled unitaries $U^{2^{k-1}}$ are then applied to the state register, controlled on the $k$th qubit of the data register, followed by performing the inverse quantum Fourier transform on the data register. Finally, the data register is measured to obtain an estimate of an eigenphase. The probability of the estimate corresponding to a particular eigenphase is given by the overlap probability of the initial state with the corresponding eigenstate.

\section{Algorithm choices}
\label{sec:alg_scaling}
In this section, we discuss the reasons for our focus on QPE and present some details of our resource estimation calculations. Further details are presented in following sections.

\subsection{Algorithm scaling}
In order to motivate our algorithmic choices, we first present simple scaling estimates for the resources required to perform the algorithms outlined in Sec.~\ref{sec:algorithms}. We present this scaling in terms of a few key parameters, ignoring the coefficients. We assume that the total time scaling takes the form
\begin{equation}
    n_{t} = \frac{n_{g} \, n_{\mathrm{rep}}}{n_{\mathrm{QPU}}}, \label{eq:timecost}
\end{equation}
where $n_{g}$ is the gate depth of a single circuit, $n_{\mathrm{rep}}$ is the number of times this circuit must be performed and $n_{\mathrm{QPU}}$ is the number of available quantum processors of suitable size, that is, the number of circuits that can be performed in parallel. The gate depth is the number of layers of gates that must be applied, where a layer of gates is a set of gates that can be applied simultaneously.

We will now outline the form of these components for the two algorithms. We define $n_{o}$ and $n_{e}$ to be the number of spin orbitals (which we note is twice the number of spatial orbitals) and number of electrons in the chemical system respectively and
\begin{equation}
    E_{\mathrm{max}} = \sum_{i=1}^{L} \vert w_{i} \vert, \label{eq:Emax}
\end{equation}
which gives the maximum possible value of any energy. We recall the $w_{i}$ coefficients are those in the qubit form of the Hamiltonian, Eq.~\eqref{eq:qubham}, and $L$ is the total number of terms. In both cases, the time scaling will also depend on the desired level of accuracy, $\epsilon$, in the energy estimate.

\subsubsection{VQE resources}
\label{sec:vqeresources}
In this section, we perform a rough estimate of the time taken to perform a VQE calculation, considering Eq.~\eqref{eq:timecost}. For a VQE calculation, the circuit depth $n_{g}$ will depend on the ansatz. There may also be some additional depth required to, for example, measure the appropriate operators; in general, we aim to make assumptions that are favourable towards VQE, and thus we will ignore this additional circuit depth here. The number of times the circuit must be repeated, $n_{\mathrm{rep}}$, will be the product of two factors -- the number of circuit applications required to obtain a single estimate of the Hamiltonian expectation, $n_{a}$, and the number of Hamiltonian expectations required in order to optimise the parameters, $n_{h}$, so that
\begin{equation}
    n_{\mathrm{rep}} = n_{a}n_{h}.
\end{equation}
A significant degree of parallelisation is possible in VQE; we will return to this once we have outlined the QPE scaling too.

We will now outline the form of the components in more detail. For a full review of  VQE and its components, see~\citeauthor{tilly2021variational}~\cite{tilly2021variational}; a scaling of VQE is also presented there, though different assumptions are made to ours.

\paragraph{Number of qubits}
We first define the number of qubits, $n_{q}$, needed to represent the relevant quantum states on the quantum computer. We will assume that we have one qubit per spin orbital, and so
\begin{equation}
    n^{\vqe}_{q} = n_{o}.
\end{equation}
This is the case for both the Jordan-Wigner and Bravyi-Kitaev transformations mentioned in Sec.~\ref{sec:qubit_hamiltonian}. It is, however, typically possible to reduce this number slightly by conserving symmetries of the chemical system~\cite{bravyi2017tapering}; however, this will not have a large effect on our calculation and so we ignore this possibility here.

\paragraph{Number of parameters}
It will next be important to consider the ansatz. The choice of ansatz plays a key role in determining the performance of a VQE calculation. The ideal ansatz would:
\begin{itemize}
    \item enable preparation of a state close to the true ground state;
    \item require as few parameters as possible, so as to minimise the time required to perform the classical optimisation;
    \item use as few quantum computational resources as possible.
\end{itemize}
In general, the ansatz circuit depth will thus depend on the accuracy, $\epsilon$, as a deeper circuit will typically allow a state closer to the true ground state to be prepared. However, it is difficult to quantify the relationship between $n_{g}$ and $\epsilon$. In this section, we will consider a fixed ansatz---the unitary coupled cluster singles-doubles (UCCSD) ansatz~\cite{romero2018strategies}. This is a chemically-inspired ansatz, which means we have reason to believe the ansatz space contains chemically-relevant systems. The circuit for this ansatz prepares the states
\begin{equation}
    \vert \psi(\boldsymbol{\theta})\rangle = e^{\hat T - \hat T^{\dagger}} \vert \mathrm{HF} \rangle,
\end{equation}
where $\vert \mathrm{HF} \rangle$ is the Hartree-Fock state and
\begin{equation}
    \hat T = \hat T_{1} + \hat T_{2} = \sum_{i,a} \theta^{i}_{a} a_{a}^{\dagger}a_{i} + \sum_{i,j, a,b} \theta^{ij}_{ab} a_{a}^{\dagger}a_{b}^{\dagger}a_{j}a_{i}.
\end{equation}
By varying the parameters $\theta_{a}^{i}$ and $\theta_{ab}^{ij}$, we can produce different quantum states. Here, $i, j, \dots$ refer to occupied orbitals in the Hartree-Fock state, and $a, b, \dots$ refer to virtual orbitals. In the VQE process, the $\theta$ parameters are optimised. We can see that the operator $\hat T$ contains components corresponding to single and double excitations of electrons from the Hartree-Fock state. For strongly-correlated systems, UCCSD may not be able to prepare a state which is suitably close to the ground state due to the limitation on the excitations considered.

The key property of the ansatz that affects the VQE calculation time is the number of parameters, $n_{p}$. For the UCCSD ansatz, this is
\begin{equation}
    n_{p} \sim n_{e}^{2}(n_{o}-n_{e})^{2},
\end{equation}
which is the scaling of the number of $\theta^{ij}_{ab}$ parameters. We will assume that both $n_{e}$ and $n_{o}-n_{e}$ scale linearly with $n_{o}$ and so
\begin{equation}
    n_{p} \sim n_{o}^{4}.
\end{equation}

Finding good ansatze is a topic of ongoing research. More recently proposed ansatz methods, which typically seek to reduce the number of parameters and/or gate depth required for a given level of accuracy, include the k-UpCCGSD~\cite{lee2018generalized} ansatz and adaptive ansatz procedures such as ADAPT-VQE~\cite{grimsley_adaptive_2019}. These have been shown to outperform UCC in some cases; however, their performance with larger chemical systems is difficult to predict. Ansatze with further improved behaviour may be developed in the future.

\paragraph{Number of Hamiltonian expectations}
The number of Hamiltonian expectations required in a particular VQE calculation is difficult to know in advance as it will depend on the shape of the ansatz parameter space. Here, we make a favourable assumption. We will assume that the number of Hamiltonian expectations required, $n_{h}$, is simply given by
\begin{equation}
    n_{h} = n_{p}.
\end{equation}
This, for example, could arise if the optimiser need only look in each parameter direction once, perhaps to verify that a minimum has already been found. Needing any fewer evaluations would imply that it was known before the calculations occurred that some parameters were not needed in the ansatz. Typical calculations will require more evaluations than this.

\paragraph{Number of ansatz circuit applications}
As mentioned above, the ansatz circuit must typically be applied many times in order to obtain an expectation of the Hamiltonian with respect to a particular state to a sufficient degree of accuracy. The number of applications needed depends on the form of the Hamiltonian and the particular quantum state.

It is typically not possible to obtain measurements of the Hamiltonian directly. However, measurements of Pauli operators can easily be obtained, and so we can make use of Eq.~\eqref{eq:qubham} in calculating the expectation of the Hamiltonian. Assuming measurements of each Pauli are obtained separately, the number of times the ansatz circuit must be performed is given by~\cite{wecker2015progress}
\begin{equation}
    n_{a} = \left ( \frac{1}{\epsilon} \sum_{i=1}^{L} \vert w_{i} \vert \sqrt{\mathrm{Var}[P_{i}]} \right )^{2},
\end{equation}
where $\epsilon$ is the desired error in the expectation estimate and
\begin{equation}
    \mathrm{Var}[P_{i}] = 1 - \langle \psi(\boldsymbol{\theta}) \vert P_{i} \vert \psi(\boldsymbol{\theta}) \rangle^{2}.
\end{equation}
The maximum value of each variance is 1 and so
\begin{equation}
    n_{a} \leq \left ( \frac{E_{\mathrm{max}}}{\epsilon} \right )^{2}.
\end{equation}
In the following we take the equality in this expression. This is likely an overestimate in practice, but we believe is sufficient to demonstrate the challenges faced. We later take a generous assumption in the scaling of $E_{\mathrm{max}}$.

We note that it is possible to reduce this through several different methods. It is possible to improve upon the assumption that we measure each Pauli individually by, for example, measuring commuting Paulis simultaneously~\cite{mcclean_theory_2016,yen2020measuring,gokhale_minimizing_2019,crawford2021efficient} or factorising the two-electron integral tensor~\cite{huggins_efficient_2019}. Such methods reduce the overall number of measurements required whilst retaining the scaling in $\left ( \frac{1}{\epsilon^{2}} \right )$. This scaling can also be improved using QPE-inspired methods at the cost of an increased circuit depth~\cite{wang_accelerated_2019, wang_minimizing_2021}; however, such increased depths are unlikely to be possible in the noisy intermediate-scale quantum (NISQ) era, before error correction is available.

\paragraph{Circuit depth}
We will assume that it is possible to perform $O(n_{q})$ parameters per layer of gates and so
\begin{equation}
    n_{g}^{\vqe} \sim n_{o}^{3}
\end{equation}
for the UCCSD ansatz outlined above.

\paragraph{Summary}
We therefore see that, for VQE, given the assumptions we have made,
\begin{equation}
    n_{q}^{\vqe} \sim n_{o}, \quad n_{g}^{\vqe} \sim n_{o}^{3}, \quad n_{\mathrm{rep}}^{\vqe} \sim \frac{n_{o}^{4} E_{\mathrm{max}}^{2}}{\epsilon^{2}}. \label{eq:vqesumm}
\end{equation}
The degree of parallelisation will depend on the total number of qubits available; we will discuss this once we have considered QPE.

\subsubsection{QPE resources}
In contrast to VQE, it is possible to make a good estimation of the quantum computational resources required to perform a QPE calculation for a given chemical system. However, this does depend on the probability overlap of the initial state with the ground state, $\eta$. In this section, we present a rough scaling of the time taken to perform a QPE calculation.

\paragraph{Circuit depth and number of repetitions}
Considering first only the parameters $\epsilon$ and $\eta$, using textbook phase estimation, one expects~\cite{lin2021heisenberg}
\begin{equation}
    n_{g}^{\qpe} \sim \frac{1}{\epsilon \eta}, \quad n^{\qpe}_{\mathrm{rep}} \sim \frac{1}{\eta}.
\end{equation}
The scaling with the properties of the Hamiltonian depends on the specifics of the quantum phase estimation calculation (see later sections for details). Using the most recent methods~\cite{lee_even_2020}, the circuit depth depends on $E_{\mathrm{max}}$, which we recall was defined in Eq.~\eqref{eq:Emax}, and $n_{o}$, the number of orbital basis functions, so that
\begin{equation}
    n_{g}^{\qpe} \sim \frac{E_{\mathrm{max}} n_{o}}{\epsilon \eta}.
\end{equation}

\paragraph{Number of qubits}
Like VQE, QPE requires approximately $n_{o}$ qubits to store the relevant quantum state. However, QPE typically also requires additional auxiliary qubits. Firstly, such qubits are needed to store the bits corresponding to the energy estimate. For the specific version of QPE outlined in Sec.~\ref{sec:QPE}, the number of additional bits is $\log_{2}\left( \frac{1}{\epsilon}\right )$; however, this can be reduced to just a single qubit using iterative phase estimation~\cite{kitaev1995quantum, kitaev2002classical}. Auxiliary qubits are also required for some methods of implementing the required unitary operators. For the most recent methods~\cite{lee_even_2020}, the number of additional qubits required is $\tilde{O}(n_{o})$, and so
\begin{equation}
    n_{q} \sim n_{o}.
\end{equation}

\paragraph{Error correction overhead}
As the circuits used when performing QPE are very deep, we expect error correction procedures to be required in order to obtain useful results from the calculations. This introduces an overhead, both in the number of qubits (spatial overhead) and the depth of the circuit (temporal overhead). We write these overheads as $\theta_{S}$ and $\theta_{T}$ respectively. For the surface code, the overheads are determined by the code distance, $d$, with $\theta_{S} \sim d^{2}$ and $\theta_{T} \sim d$. We can therefore write $\theta_{S} \sim \theta_{T}^{2}$. We note that, in order to maintain a constant probability of a logical error occurring, these overheads must increase with increasing logical circuit depth and number of logical qubits; however, they increase logarithmically and so we ignore this here. This can be seen from Eq.~\ref{eq:distance}, as will be motivated in Sec.~\ref{sec:error-corrected resource estimation}. We therefore write
\begin{equation}
    n_{q} \sim n_{o} \theta_{T}^{2}, \quad n_{g}^{\qpe} \sim \frac{E_{\mathrm{max}} n_{o} \theta_{T}}{\epsilon \eta}.
\end{equation}
For further information about the error correction overhead in the context of the surface code, see Sec.~\ref{sec:errorcorrection}.

\paragraph{Summary}
For QPE, we therefore have
\begin{equation}
    n_{q}^{\qpe} \sim n_{o} \theta_{T}^{2}, \quad n_{g}^{\qpe} \sim \frac{E_{\mathrm{max}} n_{o} \theta_{T}}{\epsilon \eta}, \quad n_{rep}^{\qpe} \sim \frac{1}{\eta}. \label{eq:qpesumm}
\end{equation}

\subsubsection{Comparison and Discussion}
We presented the scaling of the number of qubits, circuit depth and number of circuit repetitions for VQE and QPE in Eqs.~\eqref{eq:vqesumm} and \eqref{eq:qpesumm} respectively. We will now consider $n_{\mathrm{QPU}}$ in the two cases. We will assume that the total number of available qubits in the two cases is $n_{q}^{\qpe}$. We note that, should additional qubits be available, some degree of parallelisation is possible for QPE as the procedure must be repeated some number of times. However, as this factor would be the same in both the VQE and QPE scalings, we do not consider it further. Therefore, the degree of parallelisation possible for VQE is
\begin{equation}
    n_{QPU}^{\vqe} = \frac{n_{q}^{\qpe}}{n_{q}^{\vqe}} \sim \theta_{T}^{2}.
\end{equation}
Putting everything together, we therefore have, from Eq.~\eqref{eq:timecost},
\begin{equation}
    n_{t}^{\vqe} \sim \frac{n_{o}^{7} E_{\mathrm{max}}^{2}}{\theta_{T}^{2}\epsilon^{2}}, \quad n_{t}^{\qpe} \sim \frac{E_{\mathrm{max}} n_{o} \theta_{T}}{\epsilon \eta^{2}},
\end{equation}
and so
\begin{equation}
    \frac{n_{t}^{\vqe}}{n_{t}^{\qpe}} \sim \frac{n_{o}^{6}E_{\mathrm{max}} \eta^{2}}{\theta_{T}^{3}\epsilon}.
\end{equation}
We will now consider the scaling of some of these terms further. The scaling of $E_{\mathrm{max}}$ with $n_{o}$ is typically examined numerically and depends on the specific chemical system and, for example, whether the increase in $n_{o}$ is due to the increasing number of atoms or increasing basis set size. Best estimates find it scales between $n_{o}$ and $n_{o}^{3}$~\cite{lee_even_2020}. Here we take it to scale as $n_{o}$, so as to be favourable towards the VQE scaling. We can consider $\epsilon$ and $\eta$ to be constant. The degree of accuracy in the final energy estimate, $\epsilon$, will typically be taken to be chemical accuracy and thus not depend on the size of the system. Whilst $\eta$, the overlap of the initial state with the true ground state, will be system dependent, it has been suggested that it is typically possible to prepare a simple state with a good degree of overlap \cite{tubman2018postponing}. We argued above that the increase in $\theta_T$ should be logarithmic, and do not consider such logarithmic factors. We therefore find
\begin{equation}
    \frac{n_{t}^{\vqe}}{n_{t}^{\qpe}} \sim n_{o}^{7},
\end{equation}
and thus expect QPE to become preferable once the system is large enough, this size being determined by the constant in front of the scaling, which we have ignored in our analysis. This motivates our choice of QPE for the remainder of this work.

Our scaling analysis is not intended to be definitive and we acknowledge that there may be possible improvements to VQE. However, VQE presents some more general difficulties. As mentioned above, the choice of ansatz is key to determining how close it is possible to get to the true ground state -- it is difficult to guarantee that the ansatz allows preparation of a state that is close to the true ground state without having a large number of parameters and circuit depth. Furthermore, even if the ansatz can describe the desired state, there is no guarantee that the optimiser will find it -- it may instead converge to a local minimum.

In this paper, we have considered only the scaling of VQE and compared it to that of QPE. Other works have performed resource estimations for VQE and find runtimes to be prohibitively large~\cite{elfving2020will,gonthier_identifying_2020, johnson2022reducing}. For example, \citeauthor{gonthier_identifying_2020}~\cite{gonthier_identifying_2020} estimate runtimes of approximately 1-100 days to perform a single expectation evaluation for systems requiring approximately 100-300 qubits.

\subsection{QPE in this work}
Having motivated our choice of QPE, we now present some details of the algorithmic choices made in this work. Further choices are discussed in the following sections.

As mentioned above, in contrast to VQE, it is possible to make a good estimation of the quantum computational resources required to perform a QPE calculation for a given chemical system. In this paper, we present results of resource calculations, given the chemical system and desired accuracy as inputs, and further allow for several algorithmic choices to be made.

After a single run of phase estimation, an energy estimate is obtained. However, this estimate may not be within the desired accuracy of the true energy. This can be for several reasons. Firstly, it is possible that the estimate is of an energy other than the desired ground-state energy. Secondly, as the true energies can typically not be represented exactly in the finite number of chosen bits, there is some probability that an estimate, even of the correct eigenstate, will not be to the desired level of accuracy. Thirdly, it is possible that our error correction procedure failed and so a logical error occurred, making any result obtained inaccurate. It is thus necessary to repeat the phase estimation procedure several times, the number depending on the overall desired probability of success. In this paper, we do not explicitly calculate the number of repetitions needed but outline one possible method for doing so in Appendix~\ref{app:qperep}. Alternative methods exist in the literature~\cite{campbell_random_2018, lin2021heisenberg}.

\section{Implementing error corrected quantum algorithms}
\label{sec:errorcorrection}

\subsection{Quantum error correction and the surface code}

The theory of error correction is vital to practical computing schemes. All physical computers are subject to noise, and this noise can cause arbitrary errors that must be detected and corrected to ensure accurate results. In classical computers an error can flip a bit `0' to a bit `1' or vice versa, which can be corrected by a variety of schemes. In quantum computing, the task of correcting errors is dramatically more challenging. Errors on quantum computers are continuous in the general case; a qubit state $\vert \psi \rangle$ can in theory be transformed to any new state $\vert \psi' \rangle$ by noise. Noise can also entangle multiple qubits. Furthermore, measuring the state of a system to directly check for errors will cause the wave function to collapse, thus losing information if not done carefully. Quantum error correction (QEC) is designed to overcome these challenges.

One might wonder if we can manage without QEC by improving the accuracy of devices further. However, as we will see later, useful quantum circuits may contain over $10^{10}$ logical gates; the error rate of each of these gates would need to be unrealistically small with iterations of current technology to perform the full circuit without error, and proceeding without QEC is not an option for large-scale quantum computing applications.

Ultimately, QEC schemes exist that can protect against arbitrary errors, provided sufficient resources are available. In practice, this is done by encoding many \emph{physical} qubits into a \emph{logical} qubit\cite{roffe_quantum_2019}. The quantum threshold theorem then states that if the error rate on the physical qubits is below a certain threshold, the error rate on the logical qubits can be made arbitrarily small \cite{shor_fault-tolerant_1996, aharonov_fault-tolerant_2008, knill_resilient_1998, kitaev_fault-tolerant_2003}. In general, the more physical qubits available, the larger the logical qubit and the better the protection that can be achieved. As such, resource estimation for future QPE calculations must carefully include the effect of QEC.

There is a wide family of techniques for QEC. Here we shall consider the surface code, which represents each logical qubit by a $d\times d$ grid of physical qubits \cite{fowler_surface_2012}. Protecting all $d^2$ of these qubits is not possible. Instead, we seek to define just a single qubit as a protected subspace. This subspace is known as the \emph{codespace}. The state of the logical qubit is forced to reside in the codespace by measuring operators known as \emph{stabilisers}. The measurement of these stabilisers allows one to check and correct errors, without destroying information encoded in the logical qubit. A further $d^2$ \emph{syndrome} qubits are present to allow efficient measurement of the stabilisers. This leads to a total of $2d^2$ physical qubits per logical qubit.

The surface code has a number of useful properties for an error correcting code: first, the physical qubits are arranged on a 2D grid and only require nearest-neighbour connectivity; and second, the surface code can tolerate a relatively high probability of errors occurring on the physical qubits. Specifically, for a probability $p$ of an error occurring on each physical qubit per operation, the probability of an error on a logical qubit is approximately $0.1(100p)^{(d+1)/2}$ \cite{litinski_game_2019}, for each given logical operation. Note that this suggests an error threshold of $1\%$, below which the error rate of the logical qubit is decreased with increasing $d$.

\subsection{Magic state factories and the QPU architecture}
\label{sec:magic_state_factories}

\begin{figure}
    \centering
    \subfloat[]{\label{fig:base_factories}\includegraphics[width=0.5\linewidth]{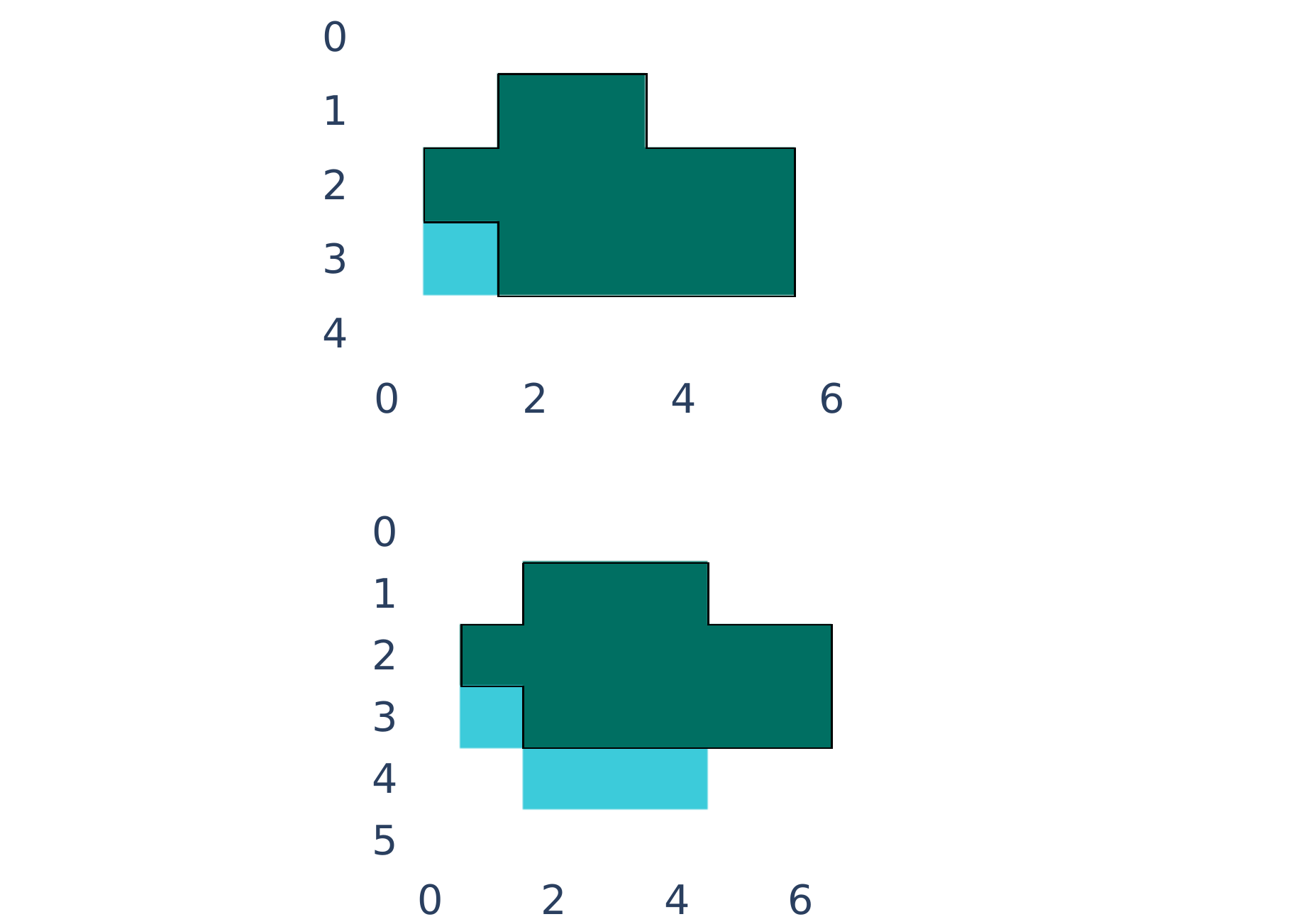}}
    \subfloat[]{\label{fig:225-to-1}\includegraphics[width=0.5\linewidth]{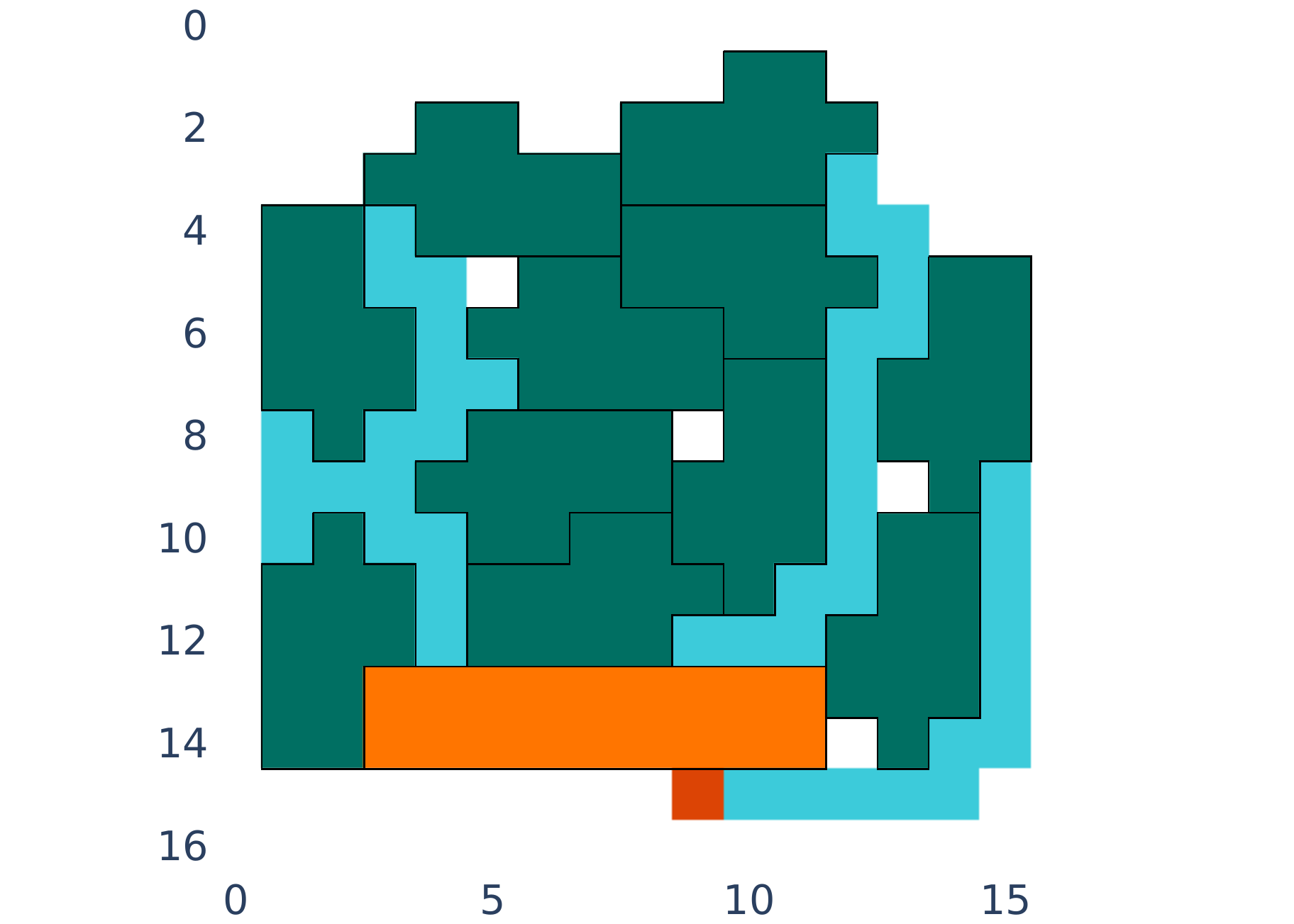}}
    \caption{\ref{fig:base_factories}: Layouts for 15-to-1 (top) and 20-to-4 (bottom) magic state factories. These consist of eleven and fourteen logical qubits respectively (green). The magic states produced are stored in the blue spaces. \ref{fig:225-to-1}: A factory which distills 225 imperfect magic states to one higher quality magic state. Eleven first-level 15-to-1 factories (green) are used to produce fifteen refined magic states, which are in turn used by the second-level 15-to-1 factory (orange) to produce one magic state of even higher quality (red). Blue lines are used to store and transport lower-quality magic states. White spaces are unused logical qubits.}
    \label{fig:factory}
\end{figure}

One challenge in QEC schemes is the Eastin-Knill theorem, which says that no QEC code can trivially implement a universal gate set \cite{eastin_restrictions_2009}. For example, the surface code can only encode Clifford gates, a collection of quantum gates which implement elements of the Clifford group. The Clifford group can be defined as the set of operations that map Paulis to other Paulis, and can be generated by the $H$, $S$ and CNOT gates, as defined in Eqs.~\ref{eq:h_s_t_gates} and \ref{eq:cnot_gate}. In order to achieve universal quantum computation we need an extra non-Clifford operation. In the surface code this is often taken to be the $T$ gate, also defined in Eq.~\ref{eq:h_s_t_gates}. A $T$ gate can be performed outside of the surface code by generating and consuming a specific quantum state, known as a \emph{magic state}. Circuits to create high-fidelity magic states are known as magic state factories \cite{litinski_game_2019, litinski_magic_2019}, and the process of creating these states is known as \emph{magic state distillation}. This process works by taking some number of noisy magic states as input and producing a smaller number of magic states, which are of higher quality, as output. The number of input and output states, the probability of success and the time taken for distillation all vary depending on the choice of factory. For example, the 15-to-1 factory from \cite{litinski_game_2019} uses 11 logical qubits, takes 15 magic states as input, and after 11 time steps produces a single magic state. Here, a time step corresponds to a single error-corrected logical operation. If the input magic states have probability $p$ of error then the probability of the distilled magic state failing is $35p^3$. In comparison, the 20-to-4 factory from \cite{litinski_game_2019} uses 14 logical qubits, runs in 17 time steps, and has probability $22p^2$ of any one of the four output magic states not being a magic state. Fig.~\ref{fig:base_factories} shows these two factories for comparison. There are also larger factories, such as the 116-to-12 factory in \cite{Haah2018codesprotocols, litinski_game_2019}, which uses 81 tiles, and produces twelve magic states after 50 time steps, such that the probability of a failed state is $41.25p^4$.

These factories can be concatenated to create even higher quality magic states, such that the magic states produced from a lower-level factory are used as input for a higher-level factory. In Fig.~\ref{fig:225-to-1} we show how the magic states produced by eleven 15-to-1 factories can be used as input for another 15-to-1 factory. This produces a 225-to-1 factory, which uses significantly more magic states and logical qubits, but in 15 time steps produces a magic state with failure probability $35(35p^3)^3 = 1500625p^9$. It is through these techniques that we can design factories which produce magic states with an arbitrarily small probability of failure.

Once a magic state is created we need to ensure that it can be routed to the logical qubits which are involved in the logical quantum circuit, which we shall refer to as \emph{data} qubits. Litinski describes a layout called the fast block, where data qubits are arranged in a 2D grid with additional \emph{auxiliary} qubits between each row of data qubits\cite{litinski_game_2019}. This arrangement allows for one magic state to be consumed by any data qubits within one time step. A time step is $d$ code cycles, where a code cycle is the time required to measure all stabilisers. Magic state factories can then be arranged around this data block to form an architecture for our quantum computer. Note that we want to design this architecture in such a way that magic states produced from each factory can reach the data block, and at the same time try to minimise the number of additional unused logical qubits -- that is, logical qubits which are not being used for data, generating magic states or routing, yet exist on the 2D grid.

\subsection{Error-corrected resource estimation}
\label{sec:error-corrected resource estimation}

We are now ready to explain how to calculate the overhead of QEC, using techniques described by Litinski \cite{litinski_game_2019}. As explained in that reference, the execution of quantum circuits can be reduced to just the application of $T$-like gates and measurements, by commuting the Clifford gates past all $T$ gates and the measurements at the end of the circuit. Thus, we focus first on magic state distillation. The number of magic states to be distilled is equal to the number of $T$ gates to be performed, denoted $N_T$. This number depends on the details of the algorithm used and will be discussed in detail in Sec.~\ref{sec:trot_vs_qub}. If we wish to achieve a total failure probability of $P_{\textrm{df}}$ for distillation, then the failure rate for each individual distillation should be no greater than $P_{\textrm{df}} / N_T$. We therefore choose a magic state factory whose failure rate satisfies this requirement. Several possible factories have been described, such as by Litinski\cite{litinski_magic_2019, litinski_game_2019} and by Haah and Hastings\cite{Haah2018}. Denote the failure rate of a particular factory by $q$. Then we choose the factory with the largest $q$ that satisfies $q \le P_{\textrm{df}} / N_T$.

We next decide the size of the fast block needed. As described above, this is the region of the quantum computer where algorithmic operations are performed on the data qubits. The data qubits are interspersed with auxiliary qubits. For $n$ data qubits, the fast block uses approximately $2n+\sqrt{8n}+1$ logical qubits in total. If $\sqrt{8n}$ is an integer then this number of qubits is exact, and the fast block is exactly square. If $\sqrt{8n}$ is not an integer then additional qubits are added or removed to the final column, as needed.

We now consider how many magic state factories are needed. The fast block can consume one magic state per time step. We therefore choose the number of magic state factories such that one magic state can be produced per time step on average. For example: the 15-to-1 factory produces a single magic state every $11$ time steps, and so we would include $11$ such factories in our setup; the 116-to-12 factory produces twelve factories every fifty timesteps and would require five factories; and the 225-to-1 factory produces a single magic state every 15 time steps, so we would require 15 factories.

Next we discuss how to arrange the magic state factories around the data block. Our aim is to arrange all factories around the data block such that each factory is connected to the data block and that the number of unused logical qubits is minimised. Problems of this nature are commonly referred to as 2D packing problems, many variants of which are NP-Hard to solve \cite{LODI2002241}, and therefore it is unlikely that an optimal solution can be found efficiently. Instead, we use a greedy algorithm, which uses a heuristic to place each individual factory in a reasonable spot based on the arrangement of the ones before it. Thus for each factory we look at every position we could place the factory, and check which ones have a path connecting the factory to the data block. We then choose the best placement for this factory based on which position minimises our heuristic. Once all factories have been placed the algorithm is complete. A pseudocode description of the algorithm is shown in Algorithm~\ref{alg:magic-arrangement}.

A key question is the choice of heuristic we optimise each placement over. One option is to minimise the number of additional logical qubits. However, there are many placements which might lead to the same number of additional logical qubits, by placing the factory around the edge of the current arrangement. Furthermore, this can lead to awkward arrangements around a data block, with a lot of wasted qubits which would be hard to use with other computations happening in parallel. Instead, we aim to minimise the perimeter of the arrangement. This heuristic ensures the arrangement remains relatively well packed by minimising gaps between factories. We also use a second heuristic to ensure that the arrangements form a roughly rectangular shape so that other computations can be more easily arranged around it. Example layouts created by this scheme are presented in Sec.~\ref{sec:results}, in Fig.~\ref{fig:arrangement}.

\begin{algorithm}
    \caption{A greedy algorithm for allocating magic state factories around a data block.}
    \label{alg:magic-arrangement}
    \begin{algorithmic}
    \Require $DataBlockShape$: Shape of the data block
    \Require $FactoryShape$ Shape of the magic state factories
    \Require $NumFactoriesTotal$: Number of magic state factories to place
    \Ensure Arrangement of magic state factories around data block
\\
        \State $NumFactoriesPlaced \leftarrow 0$
        \State $Arrangement \leftarrow DataBlock$
        \While{$NumFactoriesPlaced < NumFactoriesTotal$}
            \State $BestPlacementFound \leftarrow \emptyset$
            \State $BestPlacementCost \leftarrow \infty$
            \ForAll{$Placement \in \textrm{NewPlacements}(Arrangement, FactoryShape)$}
                \State $Path \leftarrow \textrm{FindShortestPath}(Placement, DataBlock)$
                \If{$Path \neq \emptyset$}
                    \State $PlacementCost \leftarrow \textrm{CalculatePerimeter}(Placement)$
                    \If{$PlacementCost < BestPlacementCost$}
                        \State $BestPlacementFound \leftarrow Placement$
                        \State $BestPlacementCost \leftarrow PlacementCost$
                    \EndIf
                \EndIf
            \EndFor
            \State $Arrangement \leftarrow Arrangement \cup BestPlacementFound$
            \State $NumFactoriesPlaced \leftarrow NumFactoriesPlaced + 1$
        \EndWhile
        \State \textbf{return} $Arrangement$
    \end{algorithmic}
\end{algorithm}

Finally, we choose the surface code distance, $d$. As noted above, in the surface code the error rate per logical qubit per code cycle is approximately $0.1(100p)^{(d+1)/2}$. There are $d$ code cycles per time step, and the fast block consumes an average of one magic state per time step. The total number of magic states to consume is $N_T$, and the total number of logical qubits is $N_L$. Thus, for an overall target failure rate of $P_{\mathrm{target}}$, we require that
\begin{equation}
    N_T \times N_L \times d \times 0.1(100p)^{(d+1)/2} \le P_{\mathrm{target}}.
    \label{eq:distance}
\end{equation}
Solving this equation for $d$ gives us the required surface code distance. This allows the total number of physical qubits to be calculated, as each logical qubit consists of $2d^2$ physical qubits.

Since, following Litinski \cite{litinski_game_2019}, we have reduced to quantum circuit to just the application of $T$ gates, the total runtime can be estimated as $N_T \times d \times t$, where $t$ is the time to perform one code cycle, and $d$ code cycles are performed per time step.

For resource estimates in this paper, we set the distillation failure probability as $P_{\mathrm{df}} = 1 \times 10^{-3}$ and the surface code failure probability as $P_{\mathrm{target}} = 9 \times 10^{-3}$.

\section{Trotterisation vs Qubitisation}
\label{sec:trot_vs_qub}

\subsection{Trotterisation}

As explained in Sec.\ref{sec:QPE}, the QPE algorithm estimates an eigenvalue of a unitary operator $U$. 
A natural choice is to take $U$ to be the evolution operator for some time $t$
\begin{equation}
    U=U(t)=e^{- i {\hat{H}}t}\,.
\end{equation}

Given a Hamiltonian ${\hat{H}}$, producing its corresponding evolution operator $U$ is generally a difficult task.  One can, at best, aim for a good approximation to $U$. When using $U$ to estimate an energy to a desired level of accuracy -- say, chemical accuracy -- it is paramount to control the error due to this approximation. This is usually referred as the problem of `Hamiltonian simulation' \cite{childs2018toward}.

The `Trotter approximation' is a widespread strategy for approximating $U$, given a Hamiltonian written as a sum of terms
\begin{equation}
    {\hat{H}}= \sum_{j=1}^L {\hat{H}}_j\,,
\label{eq:ham_decomoposition}
\end{equation}
each of which is easy to exponentiate -- that is, we can construct $e^{-i {\hat{H}}_j t}$ for all $j$.  Examples of such Hamiltonians include those found in chemistry.

The Trotter approximation divides the time $t$ into intervals of duration $\tau$, and considers a simple approximation to the evolution operator for each of these intervals:\footnote{The expression presented here is the so-called second order Trotter approximation, in which each operator ${\hat{H}}_j$ appears twice per time step. Other Trotter orders exist -- they differ in the number of times ${\hat{H}}_j$ appear per time step.}
\begin{equation}
    U(t) \approx U_{\textrm{Trot}}(t) = \left(\prod_{j=1}^L e^{-i \frac{\tau}{2} {\hat{H}}_j} \prod_{j=L}^1 e^{-i \frac{\tau}{2} {\hat{H}}_j}\right)^{t/\tau}\,,
    \label{eq:Trotter}
\end{equation}
with $t$ an integer multiple of $\tau$. 
The error in this approximation goes to zero for $\tau\rightarrow0$. However, the cost of implementing this approximation increases as $\tau$ decreases.  To do well in the trade-off of resources vs accuracy, one would like to choose the largest $\tau$ that gives the desired accuracy.

Given a Hamiltonian, rigorous error bounds for the Trotterisation of its evolution operator are available for finite $\tau$ \cite{childs2021theory}, but in practice these bounds tend to be very generous.  
For tighter error estimates, one can use an empirical law for the error inferred from small systems -- small systems are a numerical necessity when the error is estimated via exact diagonalisation of the Trotter and Hamiltonian operators (see, e.g., \cite{reiher2017elucidating, childs2018toward} for other empirical approaches to the Trotter error). There are a number of choices to make.  For starters, there are a variety of notions of error to quantify.  We choose $\epsilon_0$, the difference between the ground-state energies of the original Hamiltonian ${\hat{H}}$ and its Trotterised evolution operator $U_{\textrm{Trot}}(\tau)$ (it is apparent from Eq.~\eqref{eq:Trotter} that the energy spectrum of $U_{\textrm{Trot}}(t)$, defined via its eigenvalues $\{e^{-iE_i t}\}$, is a function of $\tau$).

\begin{figure}
    \centering
    \subfloat[]{
    \includegraphics[width=0.485\textwidth]{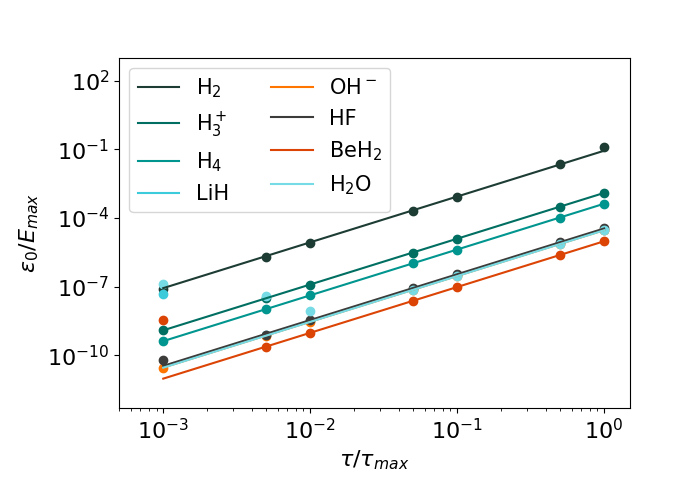}
    }
    \subfloat[]{
    \includegraphics[width=0.485\textwidth]{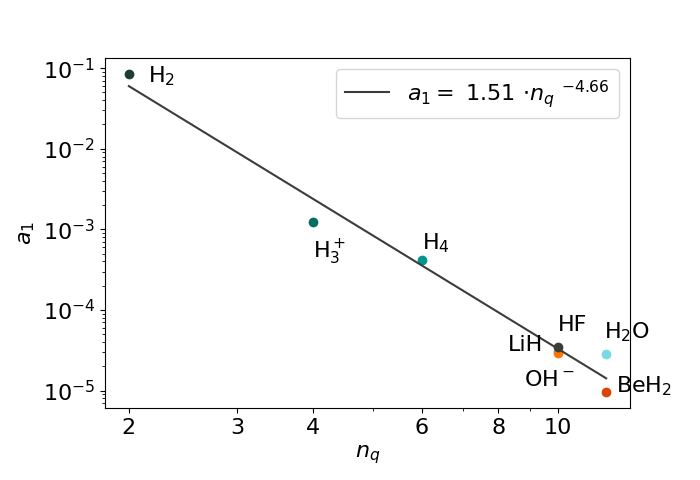}
    }
    \caption{Fit of empirical law for our set of molecules. The fit is done in two steps. In the first step (left), for each of the molecules, we generate $\delta E_0$ for $\tau/\tau_{\textrm{max}}=[1.0, 0.5, 0.1, 0.05, 0.01, 0.005, 0.001]$, and do a one-parameter fit of $\epsilon_0= E_\textrm{max}\cdot a_1 \cdot \left(\frac{\tau}{\tau_{\textrm{max}}}\right)^{2}$. Note that, for larger molecules, $\epsilon_0$ for small values of $\tau$ appear to deviate from the quadratic behaviour.  We attribute this to numerical error, and exclude these values from the fits.
    In the second step (right), we plot the $a_1$ for each molecule and fit $a_1 = a\cdot (n_q)^b$, obtaining the parameters of the empirical law in Eq.~\eqref{eq:empirical_law}: $a= 1.51\pm0.84$, $b=-4.66\pm0.27$.}
    \label{fig:Trotter_empirical_law}
\end{figure}

We inferred an empirical law from the difference between ground-state energies of ${\hat{H}}$ ($E_0$), and $U_{\textrm{Trot}}(\tau)$ ($E_{\textrm{Trot}}$) for a data set composed of small molecules (H$_2$, H$_3^+$, H$_4$, LiH, OH$^-$, HF, BeH$_2$, H$_2$O) in the STO-3G basis, in the symmetry-conserving Bravyi-Kitaev encoding -- ${\hat{H}}_j$ in Eq.~\eqref{eq:Trotter} being the Pauli strings of the Hamiltonian in that encoding.  For each molecule, this difference $\epsilon_0=E_{\textrm{Trot}}-E_0$ is well modelled by a quadratic monomial of $\tau$.  The coefficient of this monomial depends on the size of the molecule, which we characterise by the number of logical qubits needed to represent it, $n_q$. In symmetry-conserving Bravyi-Kitaev, this is two less than the number of spin-orbitals, $n_q = n_{o}-2$.  The following law results in a good fit:
\begin{equation}
    \epsilon_0= E_\textrm{max}\cdot a \cdot (n_q)^b \left(\frac{\tau}{\tau_{\textrm{max}}}\right)^{2}\,,
\label{eq:empirical_law}
\end{equation}
where $E_\textrm{max}$ is a certain bound on the maximum eigenenergy of the Hamiltonian that has been defined in Eq.~\eqref{eq:Emax}; $\tau_{\textrm{max}} \equiv \pi/E_\textrm{max}$; and $n_q$ is the number of qubits used to represent the active space of ${\hat{H}}$. 
Fig.~\ref{fig:Trotter_empirical_law} describes the fit, resulting in $a= 1.51\pm0.84$, $b=-4.66\pm0.27$.

One finds that, even with empirical laws for the error committed, the Trotter approximation is an expensive method for the Hamiltonian simulation step of QPE. For example, as quoted in Sec.~\ref{sec:results} below, for simulations of active spaces of (32e, 32o), the compilation of the Trotter operator into Clifford and $T$-gates \cite{bocharov2015efficient} gives $T$-gate counts of around $4\times 10^{14}$, for chemically accurate Trotterisation. These $T$-gate counts result in impractically long runtimes on current and projected quantum computers. We note that, whilst improvements to the Trotterisation circuits can be made~\cite{wecker2015solving}, these will not reduce the runtimes by the required orders of magnitude. More modern methods, encompassed under the names of `qubitisation' and `linear combination of unitaries' result in more moderate $T$-gate counts, of around $10^{10}$, with projected runtimes on the order of a few days.

\subsection{Qubitisation}

In the quest to find the eigenenergies of a Hamiltonian, it is actually not necessary to  solve the problem of Hamiltonian simulation and implement the time evolution operator $U=e^{-i{\hat{H}}t}$ with eigenvalues $e^{-iE_j t}$.
Instead, qubitisation methods \cite{low_hamiltonian_2019, babbush_et_al_encoding_2018} facilitate the implementation of a so-called walk operator $W$ with eigenvalues
\begin{equation}
    \operatorname{spectrum}(W) = \{ e^{\pm i \arccos(E_j/E_\text{max})} \} \label{eq:W eigenvalues}.
\end{equation}
The Hamiltonian's energies $E_j$ can then readily be retrieved by performing QPE on the walk operator $W$. The upside is that $W$ can be implemented with many fewer $T$-gates than the Trotterised time evolution $U_{\textrm{Trot}}$, for a given the Hamiltonian ${\hat{H}}$ and target accuracy.

The walk operator $W$ is simply related to the Hamiltonian ${\hat{H}}$. The starting point for the implementation of the walk operator is a decomposition of the Hamiltonian into a linear combination of unitaries (LCU):
\begin{equation}
    {\hat{H}} = \sum_{i=1}^L w_i U_i.\label{eq:LCU}
\end{equation}
The individual $U_i$ should be unitary and simple enough to be readily implementable on a quantum computer. Then the LCU decomposition can be implemented in a block-encoded fashion by using the PREPARE/SELECT framework \cite{babbush_et_al_encoding_2018,low_hamiltonian_2019}. In its most basic and simplified form, the LCU implementation is based on a state
\begin{equation}
    \vert \mathrm{PREPARE}\rangle = \frac{1}{\sqrt{E_\text{max}}}\sum_{i=1}^L \sqrt{\vert w_i\vert } \vert i\rangle,\ E_\text{max} = \sum_{i=1}^L \vert w_i\vert \label{eq:prepare}
\end{equation}
and an operator
\begin{equation}
    \mathrm{SELECT} = \sum_{i=1}^L \vert i\rangle\langle i\vert\otimes U_i \label{eq:select}
\end{equation}
which selects one of the $U_i$ based on the value of the auxiliary qubit register $\vert i\rangle$. Put together, these give
\begin{equation}
    {\hat{H}} \propto \langle \mathrm{PREPARE} \vert \mathrm{SELECT} \vert \mathrm{PREPARE} \rangle. 
\end{equation}
Any signs of $w_i$ have been absorbed into the $U_i$.
Qubitisation then shows how to construct the walk operator $W$ with the eigenvalues in Eq.~\eqref{eq:W eigenvalues} from these PREPARE and SELECT operators.

There are several flavours of qubitisation \cite{babbush_et_al_encoding_2018, berry_qubitization_2019,von_burg_quantum_2021,lee_even_2020}.
 On the one hand, by transforming or factorising the chemical Hamiltonian in different ways, they arrive at different LCU decompositions in Eq.~\eqref{eq:LCU} that promise better efficiency. On the other hand, the flavours introduce new ways to implement the PREPARE and SELECT operators, which improve upon previous approaches but can also be very tailored to their specific factorisation of the Hamiltonian.
 
Throughout, we consider the sparse method first presented in  \cite{berry_qubitization_2019} and further improved in the appendix A of \cite{lee_even_2020}. The method is tailored to the Jordan-Wigner fermion-to-qubit encoding. We select this qubitisation flavour based on its simplicity and flexible applicability to a wide range of systems.
We consider a total error $\epsilon = 1.6~\mathrm{mHa}$ of chemical accuracy \cite{lee_even_2020}. The total error is made up of three parts $\epsilon = \epsilon_{\text{TRUNC}} + \epsilon_{\text{PREP}} + \epsilon_{\text{QPE}}$ \cite{lee_even_2020}, which we chose as $\epsilon_{\text{TRUNC}}=\epsilon_\text{PREP} = 0.3~\mathrm{mHa}$ and $ \epsilon_\text{QPE}=1~\mathrm{mHa}$. In the following we explain how they arise, and which parameters of the algorithm can be adjusted to reach our total error budget of 1.6 mHa.

First of all, the chemical Hamiltonian ${\hat{H}}$ is not decomposed exactly into an LCU in Eq.~\eqref{eq:LCU}.
The sparse method exploits (approximate) sparsity in the Hamiltonian ${\hat{H}}$ by truncating small terms. We denote the truncated Hamiltonian ${\hat{H}}_\text{TRUNC}$, and perform the LCU on ${\hat{H}}_\text{TRUNC}$ instead of ${\hat{H}}$. This reduces the number of terms in the LCU decomposition, yielding a faster quantum computation. We consider two criteria to truncate the Hamiltonian according to a given error budget. First, a truncation based on the $L_2$-norm of ${\hat{H}}$. In this, the truncation threshold for the two-body coefficients is chosen such that $\vert\vert {\hat{H}} - {\hat{H}}_\text{TRUNC}\vert\vert_{L_2} \le \epsilon_{\text{TRUNC}}$ \cite{von_burg_quantum_2021}. Note that the $L_2$-norm must be taken with respect to the LCU coefficients. Then, we consider a truncation based on CCSD(T) \cite{lee_even_2020}. Specifically, we calculate the CCSD(T) energy with the original Hamiltonian ($E_{\textrm{CCSD(T)}}$) and a truncated Hamiltonian ($E_{\textrm{CCSD(T) trunc}}$) and find the largest truncation for which $\vert E_{\textrm{CCSD(T)}} - E_{\textrm{CCSD(T) trunc}} \vert \le \epsilon_{\text{TRUNC}}$. This reduces the number of terms in the Hamiltonian by up to $\sim 90\%$, which lowers the cost of implementation significantly.
 
Second, an error $\epsilon_{\text{PREP}}$  occurs when implementing the LCU decomposition with the PREPARE/SELECT machinery. The quantum circuit for PREPARE cannot implement the coefficients $\sqrt{\vert w_i\vert/E_\text{max} }$ in Eq.~\eqref{eq:prepare} to infinite precision, resulting in the rounding error $\epsilon_{\text{PREP}}$. It can be controlled by the bitlength $\aleph = \lceil \log(E_\text{max}/(2\epsilon_{\text{PREP}}))\rceil$ (A12, \cite{lee_even_2020}) for the coherent alias sampling procedure\footnote{Apart from the bitlength $\aleph$ in coherent alias sampling, the bitlength $b_r$ used in amplitude amplification of an equal superposition state in PREPARE also contributes to $\epsilon_\text{PREP}$. However the contribution of $b_r$ to the total cost is subleading and we take $b_r$ as constant \cite{lee_even_2020}.}  \cite{babbush_et_al_encoding_2018}.
The LCU can be implemented much more precisely than the evolution operator in Trotterisation;  the dependence of gate count on allowable error $\epsilon_{\text{PREP}}$ is much smaller in qubitisation. The LCU is a more efficient approximation than Trotterisation. The main reason for this is that, while Trotterisation targets the approximate implementation of $U_{\textrm{Trot}} \approx e^{-i{\hat{H}}t}$, the LCU directly targets the approximate implementation of ${\hat{H}}$, avoiding any approximation in the expansion of the exponential.

Finally, as in Trotterisation, an error $\epsilon_{\text{QPE}}$ occurs due to the final QPE, which has a finite accuracy as discussed in Sec.~\ref{sec:QPE}. Qubitisation methods typically use alternative phase estimation methods \cite{babbush_et_al_encoding_2018, luis_optimum_1996} to slightly improve on standard textbook QPE. An error $\epsilon_{\text{QPE}}$ informs that the walk operator $W$, and hence the qubitisation procedure, needs to be repeated $\lceil \pi E_\text{max} / (2 \epsilon_{\text{QPE}}) \rceil$ (eq. 44, \cite{lee_even_2020}) times in a phase estimation.

Subsequently, we will assess the time and logical qubit number needed for a quantum computer to be of aid in chemical applications. The article \cite{lee_even_2020} has taken great effort in deriving the number of logical $T$-gates and logical qubits needed for a given Hamiltonian and parameters determining the errors. The number of logical $T$-gates in the sparse qubitisation algorithm can be found by multiplying the cost (A17, \cite{lee_even_2020}) of a single iteration with the number of iterations (eq. 44, \cite{lee_even_2020}) in the phase estimation. The number of logical qubits is given by (A18, \cite{lee_even_2020}). We combine these results with error correction (see Sec.~\ref{sec:error-corrected resource estimation}) for an estimate of the physical resources required.
These results for the sparse qubitisation algorithm can be directly compared to the runtime requirements of the Trotterisation algorithm. Fig.~\ref{fig:trotter_comp} highlights the tremendous runtime advantage of qubitisation algorithms compared to Trotterisation.

\section{Drug Design Methods and the Model System}
\label{sec:chemical_system}

\subsection{Computational Chemistry in Pharma}

The interaction between drug molecules and various proteins is vital for the understanding of pharmaceutically relevant mechanisms. Unfortunately, a protein-drug system within its physiological environment may easily consist of hundreds of thousands of atoms, which makes the full quantum mechanical treatment of such systems out of reach for quantum and classical computers alike. As a consequence, the most widely used computational techniques in pharma rely on a classical (Newtonian) parametrisation of the various interactions via empirical force fields.  Current methods of rational drug design broadly belong to either ligand-based or structure-based design approaches. While the former focuses on structural features of ligands, the latter considers drug molecules within a protein environment.  Especially for the latter, an accurate description of the forces involved in protein-ligand binding is vital and the necessary force-field parameters may be obtained from quantum mechanical calculations \cite{visscher_deriving_2019,jing_polarizable_2019,nerenberg_new_2018,xu_perspective_2018}. However, while classical force fields capture most prominently the bond lengths, bond angles, dihedrals, as well as non-bonded electrostatic and van-der-Waals interactions, their traditional formulation does not account for finer electronic effects such as polarisation, charge-transfer phenomena, aromatic stacking interactions or interactions with metal ions, although extensions do exist that attempt to treat the latter phenomena with varying degrees of success \cite{sakharov2009force,cieplak2009polarization,li2017metal}. The fact that only atom types and not electrons and nuclei are considered in force-field parametrisation also renders force-field approaches incapable of describing covalent interactions and reaction mechanisms that involve bond breaking, finding transition state structures and making spectroscopic predictions. Yet despite the continued improvements in computer power and speed, the routine application of steeply scaling quantum mechanical methods in the drug design process is still very limited and mainly reserved for the study of small molecule properties and small molecule conformations.  While using semiempirical methods such as DFTB (tight-binding density functional theory) \cite{seifert2012density} and HF-3c \cite{sure2013corrected} reduces the cost significantly, these methods are often considerably less accurate than fully quantum mechanical methods. Thus, when more accurate treatment is required, embedding techniques are typically used. These methods either partition the molecule into small fragments and assemble the whole from fragment calculations, or build layers with one of them treated at a high level and the others more approximately. The great variety of these methods is reviewed elsewhere \cite{goez2018embedding,sun2016quantum}, we only remark here that some of them have been applied to protein systems containing more than 20,000 atoms \cite{ikegami2005full}. Here we are concerned with two typical choices: hybrid QM/MM and QM-cluster \cite{senn_qmmm_2009, wu_computer-aided_2021}. 

Since the ground-breaking work by Warshel and Levitt in 1976 \cite{warshel_theoretical_1976}, the idea of partitioning a chemical system into layers treated with methods of different sophistication has been a field of intense research \cite{cui_biomolecular_2021, lu_qmmm_2016}. In drug design the approach is traditionally used to describe selected residues of the binding pocket and the drug with a quantum mechanical (QM) method while the remainder of the system is simulated using molecular mechanics (MM). These hybrid QM/MM methods are generally divided into subtractive methods where the MM energy of the active site is subtracted from a sum of the QM energy of the active site and the MM energy of the entire system, and additive methods that only consider the MM energy of the environment and account for the interaction between the two systems by adding an electrostatic coupling term \cite{cao_difference_2018}. The latter describes interactions either a) solely on the MM force-field level and without any polarisation of the QM region (Mechanical Embedding), b) by incorporating point charges from the MM region in the QM Hamiltonian (Electrostatic Embedding), or c) by mutual polarisation of the regions requiring a polarisable MM force field (Polarisable Embedding) \cite{senn_qmmm_2009}. 

In a QM-cluster approach the active site is physically cut out of its environment only considering the drug and the nearest interacting amino acids. Cross sections are saturated by usually hydrogen atoms or methyl groups and constraints are added to ensure the rigidity imposed by the protein surroundings. Electrostatic effects are compensated using continuum solvation and a dielectric constant \cite{himo_recent_2017, cerqueira_protocol_2018}. Both the hybrid QM/MM and the QM-cluster method are used to gain insight in the drug-protein binding and electronic processes in the binding pocket like electronic excitations \cite{boulanger_qmmm_2018} or mechanisms of binding or action \cite{hu_accurate_2013}. However, both methods are restricted to a few hundred atoms at most depending on the level of description which is not enough to describe e.g. effects of ligand binding at other sites than the binding pocket (allosteric) or other large scale mechanisms.

Finally, it should be mentioned that a number of embedding methods have already been proposed for use with quantum computers in an attempt to reduce the heavy resource requirements. Local approaches to active space construction have been recently proposed \cite{pandharkar2022localized} and applied to quantum computing \cite{otten2022localized}. The quantum variants of dynamical mean field theory \cite{Bauer2016} and density matrix embedding theory (DMET) \cite{rubin2016hybrid} were published some years ago and costing studies are also available for DMET. \cite{yamazaki2018towards} Energy-weighted DMET\cite{tilly2021reduced} and Gutzwiller variational embedding\cite{Yao2021} approaches have been tested on current quantum processors. Pharmaceutical model systems have also been studied including a study of protein-ligand interactions using DMET \cite{kirsopp2021quantum} and our own work on a multilayer embedding approach \cite{izsak2022quantum}.

\subsection{Active Space Selection}

In both the QM/MM and QM-cluster approaches, a central QM region is selected to be treated at the highest level of theory. Unfortunately, this region is typically still too large to be treated directly on a quantum device. To construct the molecular Hamiltonian within this region, an active space of orbitals must be selected in a manner reminiscent of frozen core \cite{fock1940incomplete,huzinaga1971theory} and complete active space \cite{roos1980complete} approaches. In our previous work, we suggested a way how the active orbitals may be selected using local fragment occupied orbitals and a corresponding set of natural orbitals obtained from perturbation theory \cite{izsak2022quantum}. We also outlined a secondary subtractive embedding process to take care of correlation effects outside the active space. For the purposes of resource estimation, this second step is not necessary.

\subsection{The Model System}

As a model system for the subsequent resource estimation within a QM cluster approach, we have chosen the drug Ibrutinib which was approved for treatment of non-Hodgkin lymphoma by the FDA in 2015 \cite{us_food_and_drug_administration_fda_2015}. It inhibits Bruton's tyrosine kinase (BTK) - a vital element of the B-cell receptor signaling pathway - and thus induces apoptosis in B-cell tumours \cite{honigberg_bruton_2010}. It covalently binds to cysteine 481 in BTK via a Michael addition reaction \cite{voice_mechanism_2021}. Successful binding of a drug to a target depends on many factors in both the binding pocket and its environment. In order to design drugs efficiently we need to gain a thorough understanding of the binding process. In the first step a covalent drug binds in the same manner as a non-covalent drug, namely via weak interactions. If a reactive electrophilic group on the drug is then positioned in proximity and favourable arrangement to a nucleophilic group on the protein the covalent bond can be formed via an electronic rearrangement. The latter cannot be described by most commonly used computational drug design methods. \cite{heifetz_design_2020}

\begin{figure}
    \centering
\includegraphics[width=0.5\linewidth]{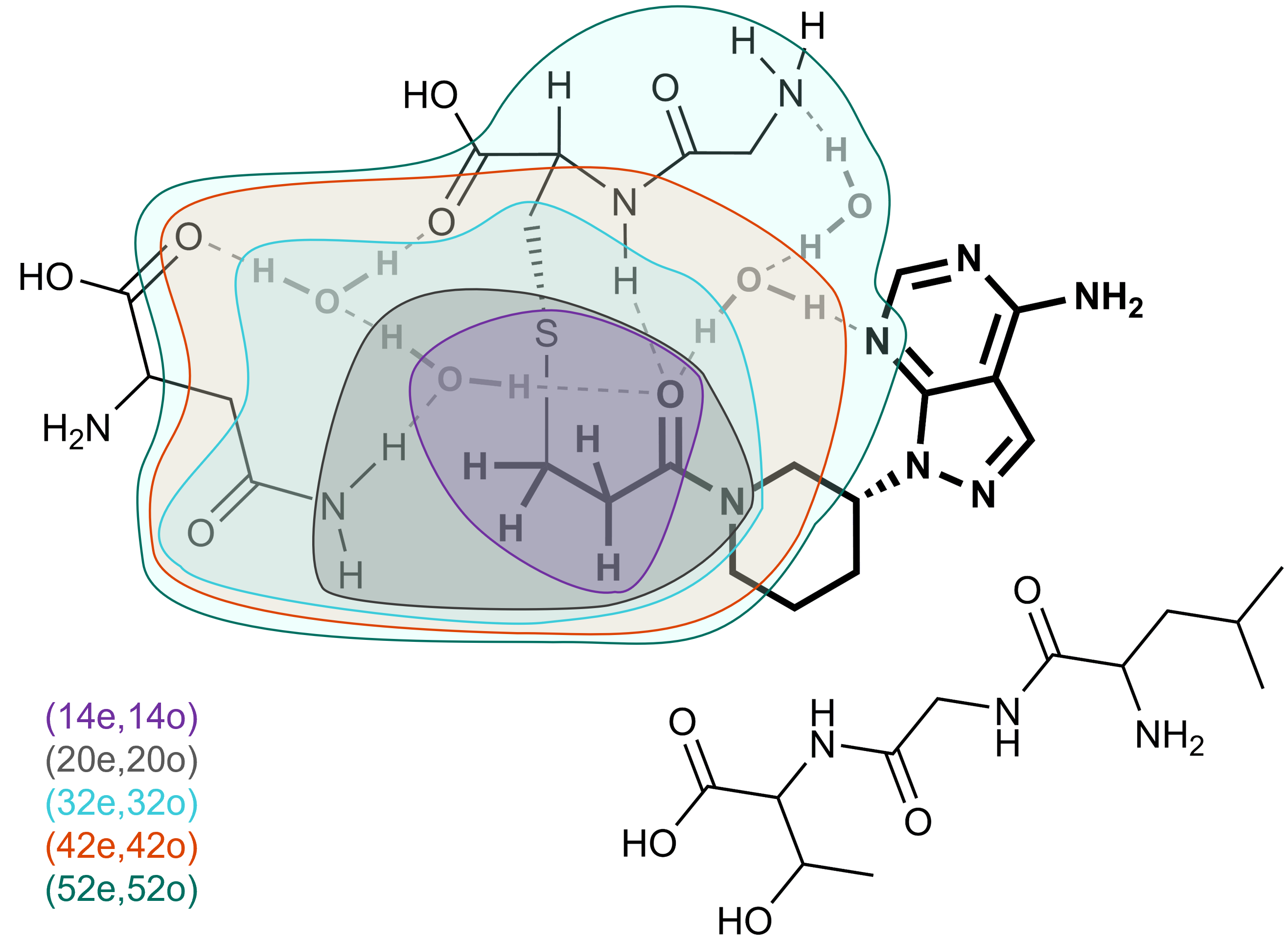}
    \caption{The cluster containing part of the binding pocket and the Ibrutinib inhibitor. The various fragments in which the active space orbitals were selected are indicated using various colors.}
    \label{fig:active}
\end{figure}

The size of the cluster has do be decided depending on the residues, ligand groups and water molecules or ions contributing to the binding or mechanism. To ensure the correct atomic arrangements and to represent the rigidity of the binding pocket it is crucial that the underlying crystal structure is well resolved and that the bonds cut and the atoms fixed are carefully chosen \cite{ahmadi_multiscale_2018}. A cluster containing the ligand, all neighbouring residues (within ~5 \AA{} of ligand) and water molecules would account for over 400 atoms and thus be too big for our purposes. Instead, a medium sized cluster was selected in which the ligand was cut beyond the pyrazolo pyrimidine moiety and which also included residues Leu408, Gly409, Thr410, Gly480, Cys481, Asn484 and four water molecules. The cluster contained 129 atoms and was fixed at position 3 of the piperidine ring (Fig.~\ref{fig:active}). There have been numerous studies in the past that have used sizes similar to the cluster-size considered here, although the treatment of a larger system would have been computationally feasible \cite{cerqueira_mechanism_2017, Prejano_qm_2018, Blomberg_structure_2020}.  The cluster represents the product structure of the binding mechanism of the formation of the covalent bond between the ligand and Cys481 of the protein \cite{voice_mechanism_2021}. In our approach, the selection of occupied orbitals corresponds to selecting fragments. We selected active fragments in five different sizes as shown in  Fig.~\ref{fig:active}. The smallest chosen area comprises the ligand warhead up to the carbonyl group, sulphur of cysteine and an $OH^{-}$ unit from the interacting water molecule (purple) and represents the minimum of the reacting and primarily interacting atoms for the boond formation step. Incrementally adding more interacting atoms from the binding pocket and larger parts of the ligand and the respective amino acid, the largest chosen area captures all relevant interactions and includes the ligand's warhead moiety up to the two closest $CH_{2}$ and the coordinating pyrazolo pyridine nitrogen atom, four water molecules, Gly480 and Cys481 excluding the saturation groups and the functional tail and carbonyl oxygen of Asn484 (green). 

\subsection{Computational Methods}

Input structures for all calculations were based on the crystal structure of a covalently bound ibrutinib/BTK complex by Bender et al. (PDB ID: 5P9J, 1.08 \AA{} resolution) \cite{bender_ability_2017}. Using Maestro's \cite{noauthor_schrodinger_nodate-1} Protein Preparation Wizard \cite{madhavi_sastry_protein_2013} missing residues were filled in using Prime \cite{jacobson_hierarchical_2004, jacobson_role_2002}, hydrogen atoms were added and refined with PROPKA \cite{sondergaard_improved_2011, olsson_propka3_2011}. The C$\alpha$  atoms were constrained for every terminal amino acid included in the cluster and  one atom in the ligand to account for their positions in the X-ray structure. Geometry optimisations were carried out with Jaguar 11.2 \cite{bochevarov_jaguar_2013} using the DFT functional B3LYP \cite{becke_densityfunctional_1993, lee_development_1988}, Grimme's dispersion correction D3 \cite{grimme_consistent_2010,grimme_effect_2011} and the 6-31G+** basis set. The CPCM (conductor-like polarizable continuum) solvation model \cite{cossi_energies_2003} with the dielectric constant $\varepsilon$ = 4 was used to describe the effect of the global protein environment \cite{siegbahn_transition-metal_2000, siegbahn_quantum_2011}. A frequency calculation was carried out to confirm that the only imaginary modes present are small and resulting from the atomic constraints, confirming the structure to be a minimum. The active space integrals were calculated using the ORCA program package \cite{ORCA}.  The def2-TZVP basis set was used \cite{weigend2005balanced}. The occupied orbitals were localized using the Pipek-Mezey approach \cite{pipek1989fast} and were mapped to fragments using intrinsic atomic orbitals \cite{knizia2013intrinsic} and the criterion that the orbital charge on the fragment be larger than 0.95 \cite{izsak2022quantum}.  The same number of active virtuals were selected using perturbation theory as the number of active occupied orbitals obtained in the previous step \cite{izsak2022quantum}.

\section{Results}
\label{sec:results}

We now present results of our resource estimations. We consider the molecule and active spaces discussed in Sec.~\ref{sec:chemical_system}, which have sizes from (14e,14o) to (100e,100o).

We perform resource estimation for the two QPE approaches described in Sec.~\ref{sec:trot_vs_qub}. In the first approach we consider the textbook QPE algorithm using Trotterisation to a precision of 1.6 mHa. We estimate the error from Trotterisation using the empirical law described in Eq.~\eqref{eq:empirical_law}. In the second approach we consider the Heisenberg-limited QPE algorithm described by Lee \emph{et al.}~\cite{lee_even_2020, babbush_et_al_encoding_2018} using qubitisation, specifically the sparse qubitisation method as described in Sec.~\ref{sec:trot_vs_qub}. The overall precision is again taken to be 1.6 mHa. We refer to these two approaches as `QPE with Trotterisation' and `QPE with sparse qubitisation' in the following, although it is important to emphasize that improvements in the latter are not solely due to the use of qubitisation.

Physical error rates of $0.01\%$ and $0.1\%$ ($p=10^{-4}$ and $p=10^{-3}$) are considered. The code cycle duration is taken to be $1$ $\mu$s, which is believed to be realistic for future superconducting quantum processors.

\begin{figure}
    \centering
    \includegraphics[width=0.8\textwidth]{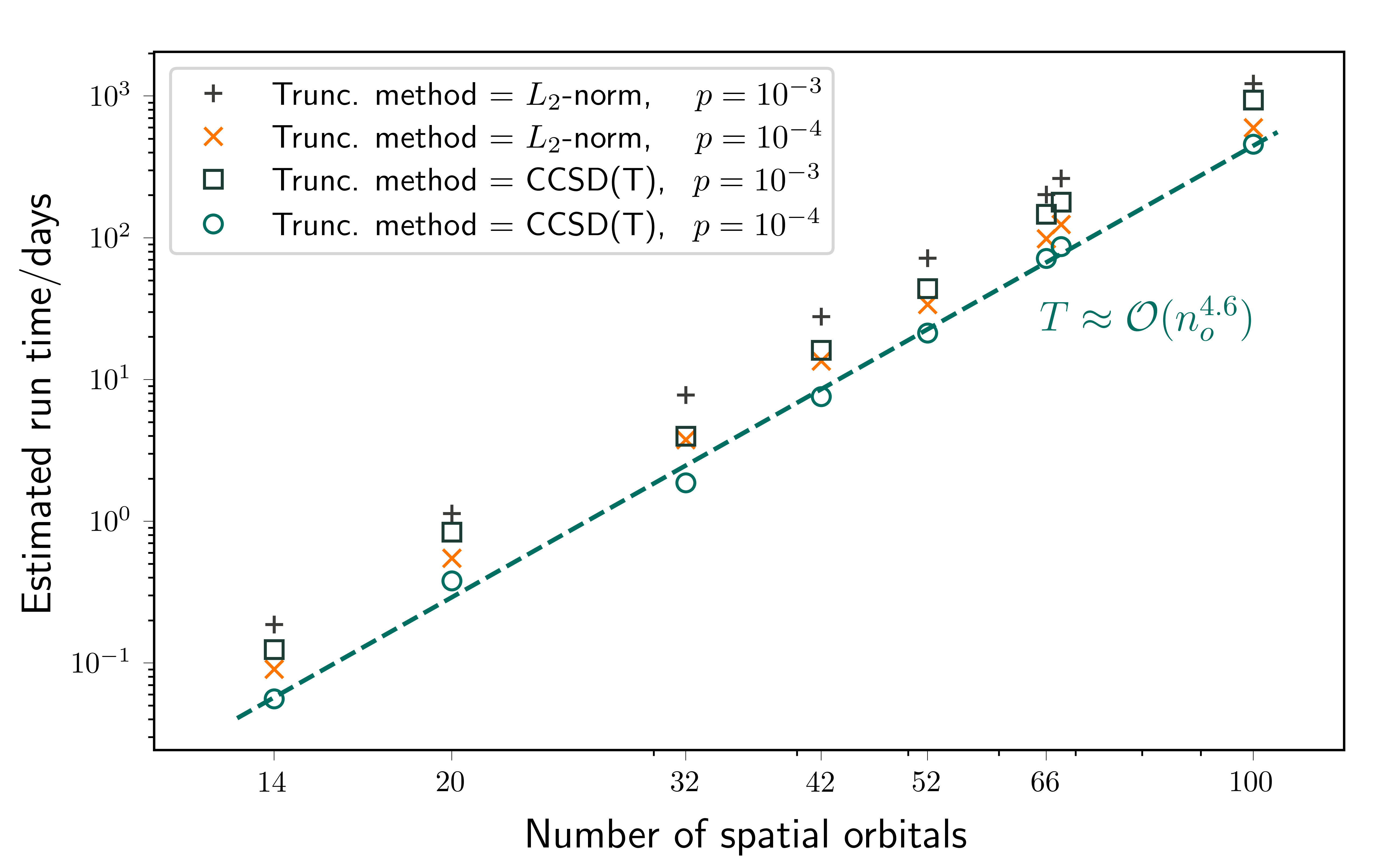}
    \caption{The runtime to perform QPE using sparse qubitisation. Active spaces from (14e,14o) to (100e,100o) are considered. It is assumed that one time step takes $1$ $\mu$s to perform. Physical error rates, $p$, of $0.01\%$ and $0.1\%$ are considered. The Hamiltonian is either truncated using an $L_2$-norm criterion or a CCSD(T) criterion. In each case, the runtime scales as approximately $n_o^{4.6}$ with the number of active orbitals.}
    \label{fig:runtime}
\end{figure}

Fig.~\ref{fig:runtime} presents the runtime for QPE with sparse qubitisation as a function of active space size, considering both error rates ($p=10^{-4}$ and $p=10^{-3}$), and both Hamiltonian truncation approaches defined in Sec.~\ref{sec:trot_vs_qub}. On this log-log plot a reasonable power law fit is evident. The runtime is found to scale as roughly $T \sim \mathcal{O}(n_o^{4.6})$, where $n_o$ is the number of spatial orbitals. The power law plotted is fit using data with $p=10^{-4}$ and CCSD(T) truncation only, but the same scaling is observed for each set of data. A power law with respect to the number of orbitals was already anticipated in \cite{berry_qubitization_2019} appendix D, and the exponent found is in approximate agreement with our observations. For the trivial (14e,14o) we estimate a runtime of 1.3 or 3.0 hours with $p=10^{-4}$ and $p=10^{-3}$, respectively (and using CCSD(T) to assess the truncation criterion). For (32e,32o) the corresponding runtimes are 1.9 or 4.0 days. For (100e,100o), we estimate respective runtimes of 1.3 and 2.6 years. These runtimes are high, but are likely to reduce with further algorithmic developments.

We also considered using both CCSD(T) and the $L_2$-norm to assess the truncation criterion. In both cases, we aimed for a Hamiltonian truncation error of $0.3$ mHa or less. As can be seen in Fig.~\ref{fig:runtime}, the CCSD(T) metric allows more Hamiltonian terms to be truncated, resulting in fewer $T$ gates overall. The number of $T$ gates is typically within a factor of $1.2$ to $2.0$ between these two approaches, for the active spaces studied here. The estimated numbers of $T$ gates in each approach are presented in Table~\ref{tab:num_t_gates}. Note that the runtime to perform CCSD(T) is negligible compared to the estimated QPE runtime for the active spaces considered. This CCSD(T) cost is not included in the presented runtimes. If QPE algorithms become efficient enough that thousands of orbitals can be treated then this situation may eventually change, in which case using the $L_2$-norm may be preferable.

\begin{figure}
    \centering
    \includegraphics[width=0.9\textwidth]{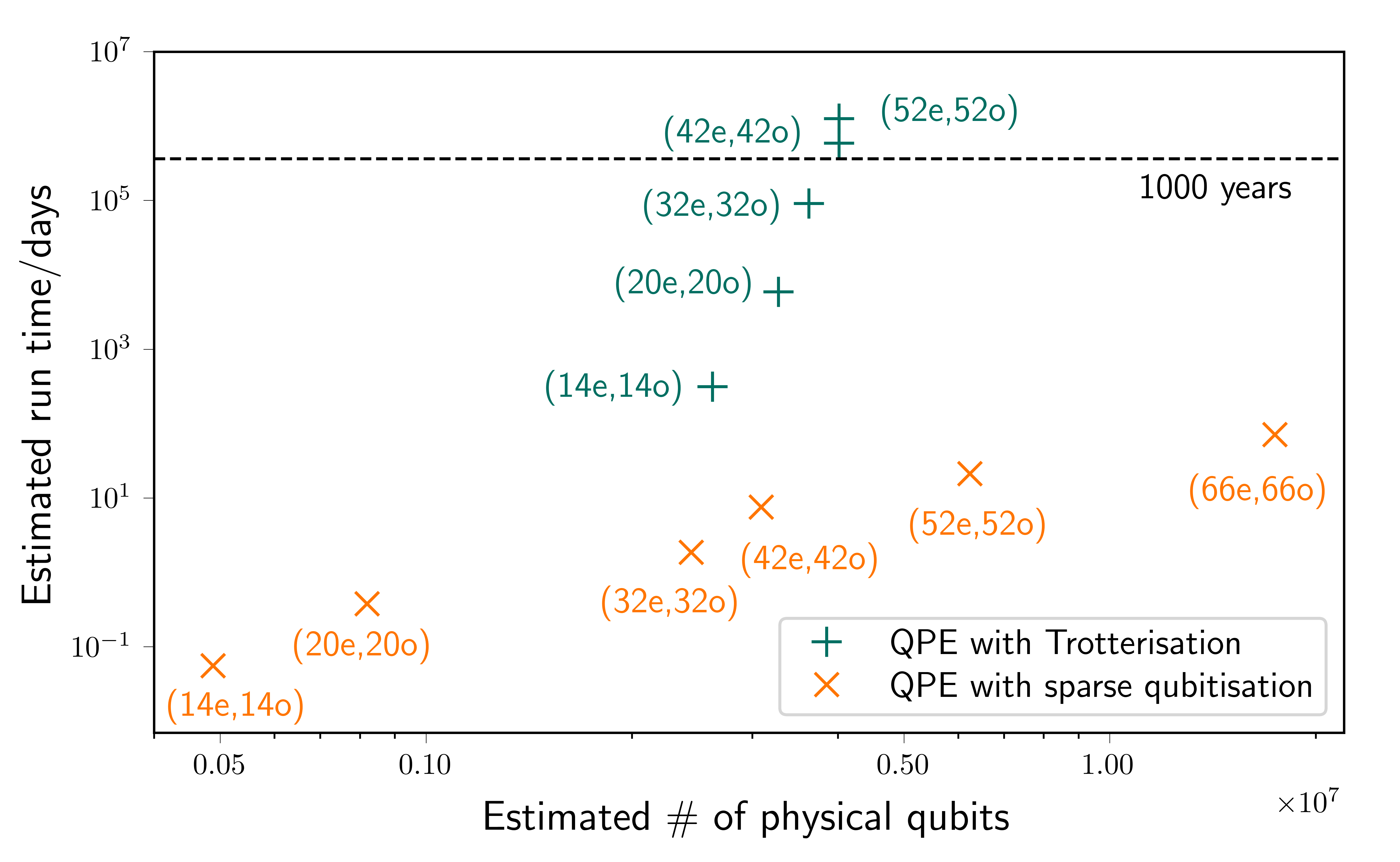}
    \caption{Comparison of resources (runtime and total number of physical qubits) using two QPE algorithms. The first (orange) used qubitisation, and the Hamiltonian was truncated to remove small terms up to an error budget. The second (green) used textbook QPE with Trotterisation and no truncation of the Hamiltonian. The latter algorithm has a much steeper scaling in runtime. Even for a (14e,14o) active space the runtime is multiple orders of magnitude more expensive.}
    \label{fig:trotter_comp}
\end{figure}

In Fig.~\ref{fig:trotter_comp} we compare the runtime and total number of physical qubits between the two QPE approaches defined above, with a physical error rate of $p=10^{-4}$. It is seen that the Trotterized approach is dramatically more expensive than the sparse qubitisation method, and has significantly steeper scaling in runtime with active space size. For example, the (32e,32o) example, which takes $1.9$ days in the latter method, is estimated to take roughly $250$ years in the Trotterised algorithm.

To calculate the number of physical qubits on the QPU, we consider the full layout of the fast block, magic state factories, and routing qubits. Making an assumption that the overall QPU will be rectangular, we then find the smallest rectangle that encloses the fast block and all magic state factories. The number of qubits within this rectangle defines the total qubit count in our results. It is seen that the total number of physical qubits is increased in the Trotterised algorithm. This is interesting as there is a significant data qubit overhead associated with performing the qubitisation algorithm. However, the increased number of $T$ gates in the Trotterised algorithm necessitates a higher surface code distance, such that the number of physical qubits is increased overall. Moreover, while the number of data qubits in the fast block is much lower when performing Trotterisation, the QPU architecture may be dominated by several large magic state factories. Note that the number of physical qubits is the same for both (42e,42o) and (52e,52o) active spaces in Fig.~\ref{fig:trotter_comp}, when using Trotterisation. This is because the same factory arrangements were used for both, and the required surface code distance is also found to be the same.

\begin{table*}
\begin{center}
{\footnotesize
\begin{tabular}{@{\extracolsep{4pt}}cccc@{}}
\hline
\hline
 & \multicolumn{2}{c}{Qubitisation} & \multicolumn{1}{c}{Trotterisation} \\
\cline{2-3} \cline{4-4}
\# of spatial orbitals & $L_2$-norm truncation & CCSD(T) truncation & No truncation \\
\hline
14   & $5.6 \times 10^8$ & $3.7 \times 10^8$ & $1.6 \times 10^{12}$ \\
20   & $3.2 \times 10^{9}$ & $2.3 \times 10^{9}$ & $2.7 \times 10^{13}$ \\
32   & $2.0 \times 10^{10}$ & $1.1 \times 10^{10}$ & $4.0 \times 10^{14}$ \\
42   & $6.9 \times 10^{10}$ & $4.1 \times 10^{10}$ & $2.4 \times 10^{15}$ \\
52   & $1.7 \times 10^{11}$ & $1.1 \times 10^{11}$ & $5.2 \times 10^{15}$ \\
66   & $4.7 \times 10^{11}$ & $3.4 \times 10^{11}$ & - \\
100  & $2.7 \times 10^{12}$ & $2.1 \times 10^{12}$ & - \\
\hline
\hline
\end{tabular}
}
\caption{The required number of $T$ gates to perform QPE for various active spaces. No truncation of the Hamiltonian is performed for QPE with Trotterisation. For QPE using qubitisation the Hamiltonian is truncated using both CCSD(T) and the $L_2$-norm to assess the error incurred, with a target truncation error of $0.3$ mHa or less. The CCSD(T) criterion truncates more terms, resulting in a lower estimate for the required number of $T$ gates.}
\label{tab:num_t_gates}
\end{center}
\end{table*}

\begin{figure}
    \centering
    \includegraphics[width=0.49\linewidth]{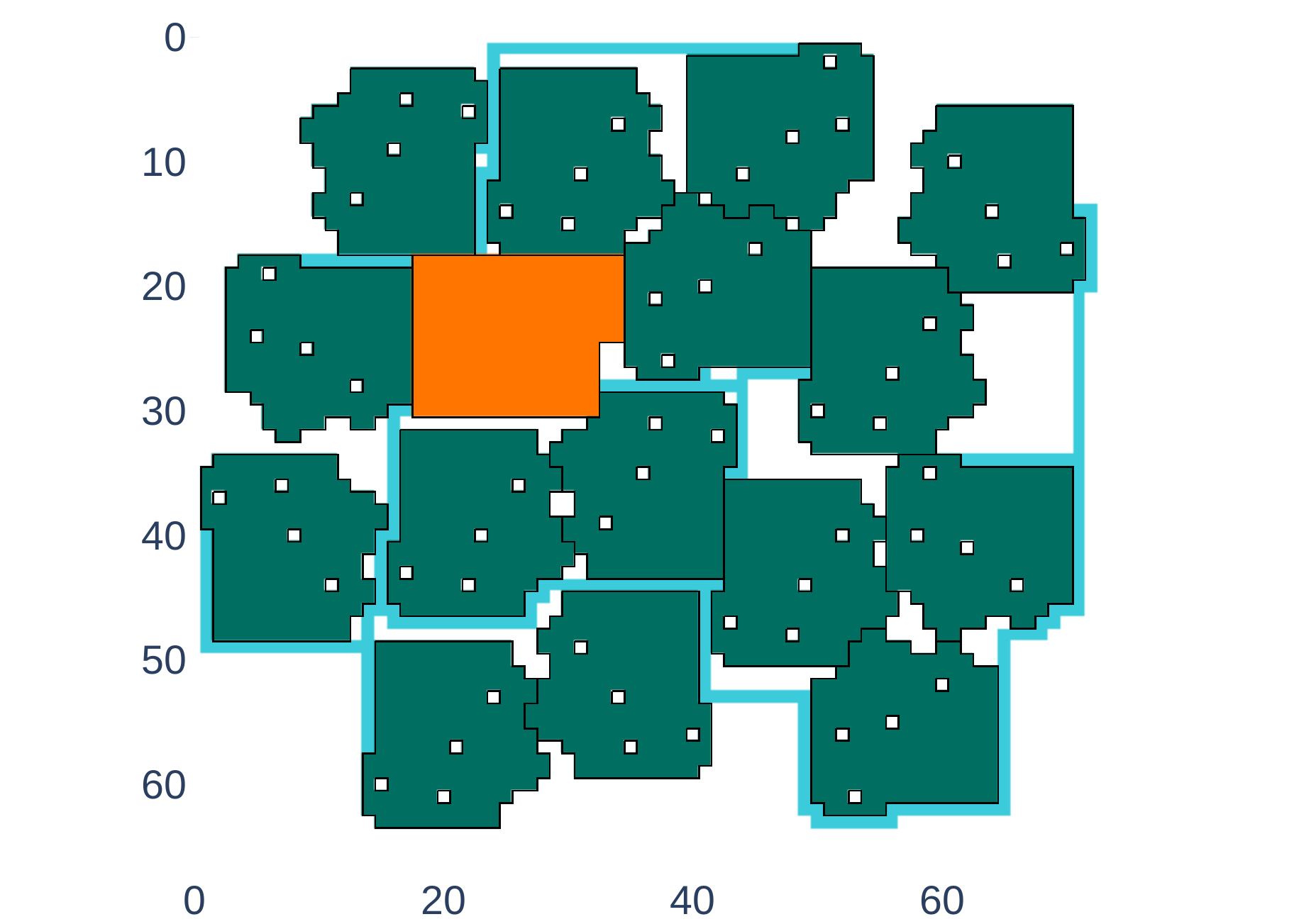}
    \includegraphics[width=0.49\linewidth]{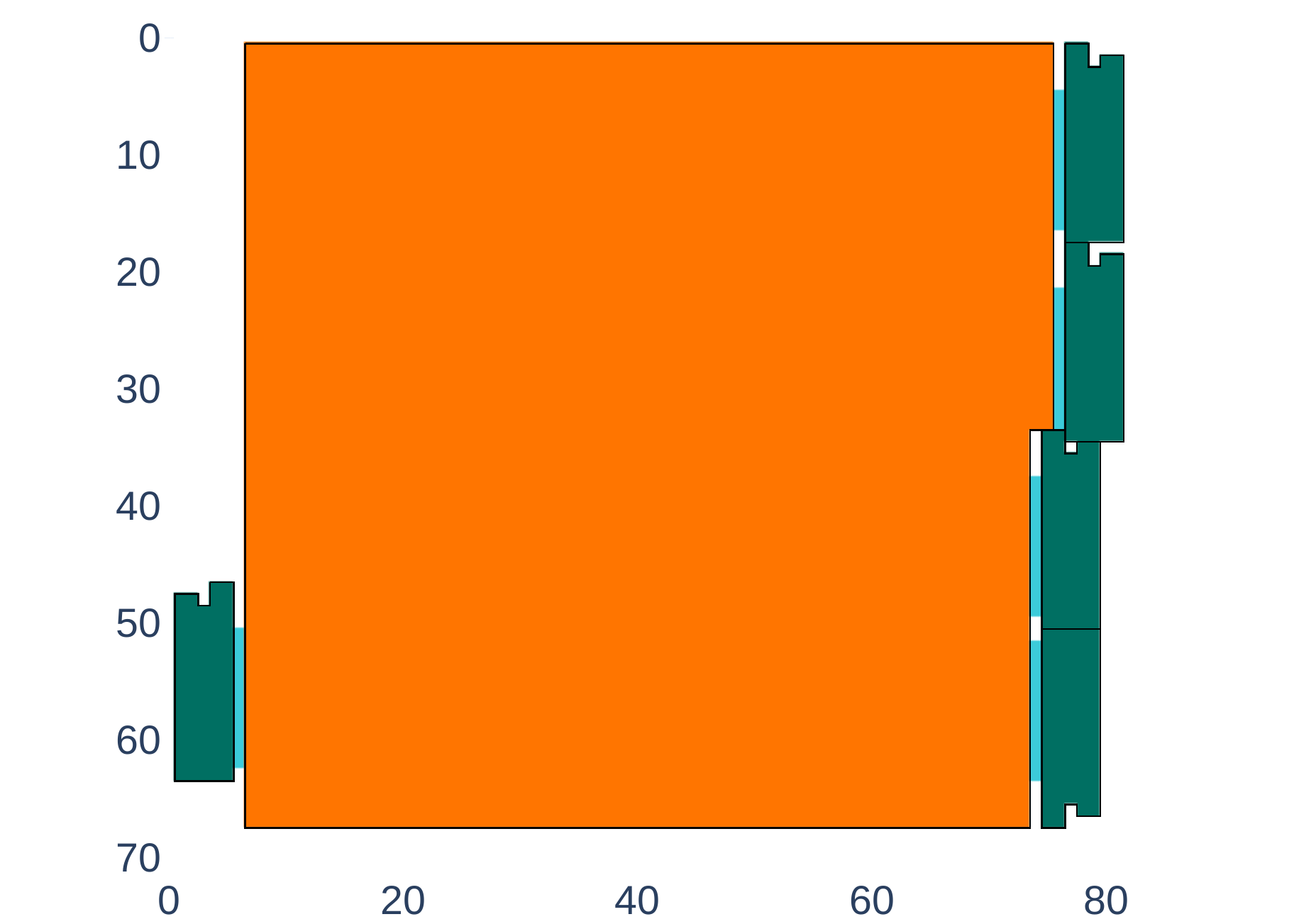}
    \caption{QPU layouts used to perform QPE experiments on the (32e,32o) active space example. Left: layout used for QPE with Trotterisation. Right: Layout used for QPE with qubitisation. Data block qubits are orange, magic state factory qubits are green, and routing and storage qubits are blue. Qubitisation uses many more data qubits such that the data block is much larger. However the higher $T$-gate count in QPE with Trotterisation necessitates larger magic state factories (225-to-1) compared to those in qubitisation (116-to-12). Axes are included to indicate the total number of logical qubits in both layouts, with each logical qubit having size 1-by-1. However, note that the code distance is higher in QPE with Trotterisation (see Table~\ref{tab:distance}) so that these are not to physical scale.}
    \label{fig:arrangement}
\end{figure}

To present a specific example in more detail, we again consider the (32e,32o) active space using the Trotterisation QPE approach. Here the number of required data qubits is $82$. The required number of $T$ gates is $N_T = 3.96 \times 10^{14}$. In order to perform magic state distillation for all such $T$ gates with the desired success probability (see Sec.~\ref{sec:error-corrected resource estimation}), we use 225-to-1 magic state factories, whose layout is presented in Fig.~\ref{fig:225-to-1}. This factory produces one magic state every fifteen time steps. Thus, we include fifteen magic state factories in order to produce one magic state per time step. Note that there may be smaller magic state factories that suffice and which we have not considered here. However, the 225-to-1 factory is optimal from those considered in this work. Using the approach presented in Algorithm~\ref{alg:magic-arrangement}, we generate a layout for the device, presented in Fig.~\ref{fig:arrangement}. The total number of logical qubits for the fast block, magic state factories and for routing is $3226$. We then solve Eq.~\ref{eq:distance} with $p=10^{-4}$, $N_L = 3226$ and $N_T$ as above, giving a required code distance of $d=20$. The smallest rectangular region which contains the above layout consists of $4536$ tiles in total. Lastly we note that there are $2d^2 = 800$ physical qubits per logical qubit. Thus, the total number of physical qubits is estimated as $4536 \times 800 = 3.6 \times 10^6$, as plotted in Fig.~\ref{fig:trotter_comp}. The total runtime is estimated as $N_T \times d \times 10^{-6}$ s $= 7.9 \times 10^9$ s.

\begin{table*}
\begin{center}
{\footnotesize
\begin{tabular}{@{\extracolsep{4pt}}ccccc@{}}
\hline
\hline
 & \multicolumn{2}{c}{Qubitisation} & \multicolumn{2}{c}{Trotterisation} \\
\cline{2-3} \cline{4-5}
\# of spatial orbitals & $p = 10^{-4}$ & $p = 10^{-3}$ & $p = 10^{-4}$ & $p = 10^{-3}$ \\
\hline
14   & 13 & 29 & 17 & 36 \\
20   & 14 & 31 & 19 & 39 \\
32   & 15 & 32 & 20 & 41 \\
42   & 16 & 34 & 21 & 43 \\
52   & 17 & 35 & 21 & 43 \\
66   & 18 & 37 & - & - \\
100  & 19 & 39 & - & - \\
\hline
\hline
\end{tabular}
}
\caption{The required surface code distances for various active spaces. We consider QPE performed using Trotterisation and the full Hamiltonian, and QPE using qubitisation and truncating small Hamiltonian elements. Physical error rates ($p$) of $10^{-4}$ and $10^{-3}$ are considered.}
\label{tab:distance}
\end{center}
\end{table*}

In the QPE approach with sparse qubitisation, the same (32e,32o) problem requires $2207$ logical qubits, but only $N_T=1.1 \times 10^{10}$ $T$ gates. In this case the 116-to-12 factory suffices. The device layout is again presented in Fig.~\ref{fig:arrangement}. A similar calculation as above leads to the lower runtime and number of physical qubits as in Fig.~\ref{fig:trotter_comp}. This dramatic reduction in runtime emphasises the importance of recent developments, and the potential value of similar developments in the future.

Lastly we investigate the required surface code distance in each case, as presented in Table~\ref{tab:distance}. The qubitisation results here used CCSD(T) as the Hamiltonian truncation criterion (using the $L_2$-norm criterion makes no significant difference to the required distance). The main factor affecting the required code distance is the physical error rate. For example, for the (32e,32o) active space the code distance increases from $d=15$ to $d=32$ as $p$ is increased from $10^{-4}$ to $10^{-3}$, in QPE with qubitisation. For a fixed $p$ the code distance is less sensitive to the $T$-gate count. For (32e,32o) the code distance increases from $d=15$ to $d=20$ between the two QPE approaches, although the number of $T$ gates increases significantly from $1.1 \times 10^{10}$ to $4.0 \times 10^{14}$. This emphasises that improvements in device fidelities can significantly reduce the challenge of performing an error-corrected algorithm in practice.

\section{Conclusions}
\label{sec:outlook}

In this paper we have presented an overview of resource estimation for quantum computing calculations in pharmaceutical applications. This has focused on quantum phase estimation (QPE), which was first introduced in the 1990's, but has recently undergone a number of significant developments to reduce its practical cost. We have also performed a detailed costing of quantum error correction (QEC) in QPE applications, particularly the surface code, which will be crucial to performing quantum computation in practical problems.

We performed QPE resource estimation for several active spaces of the drug Ibrutinib. QPE was costed using two techniques: Trotterisation with the full Hamiltonian, and qubitisation using a truncated Hamiltonian. We find a dramatic improvement with the latter technique; calculating the ground-state energy in a (42e,42o) active space is estimated to take over 1000 years using Trotterisation, which is reduced to around 7.6 days using the sparse qubitisation approach (assuming a physical error rate of $0.01\%$, and code cycle duration of $1$ $\mu$s). This emphasises that algorithmic improvements can reduce the cost of quantum computing by several orders of magnitude, and are transformative to the potential power of quantum computers. Some of the runtimes remain high; for example, obtaining the ground-state energy for a (100e,100o) active space is estimated to take over a year. This emphasises that further algorithmic improvements are important. Given the dramatic reduction in runtime seen above, we  expect such improvements to occur. For example, our costing assumed that all $T$-gates are performed in serial, whereas the runtime can be reduced in theory through parallel execution \cite{litinski_game_2019}. Further truncation of the Hamiltonian may be possible through techniques such as tensor hypercontraction \cite{lee_even_2020}. QEC is also an extremely active area of research, and improvements here may further reduce the resources required for large-scale applications. It should be noted that current quantum computers have a low qubit count compared to those presented in our resource estimates. For example, IBM's Eagle processor has 127 qubits\cite{ibm_eagle}. An experiment by Google Quantum AI has recently been performed which demonstrates decreasing logical error rate with increasing qubit number\cite{google_qec_2022}. However, the authors caution that their error rates are still close to the code threshold and must be reduced further to facilitate ``practical scaling''. Thus, the state-of-the-art is still some way from performing non-trivial QPE calculations.

In assessing the usefulness of quantum computers for pharma, several factors must be considered. In this paper, we focused on demonstrating that QPE running on fault tolerant quantum computers will be able to handle large active spaces. It remains important to ensure that the accurate treatment of this quantum region is coupled with a balanced treatment of the environment lest the errors coming from the latter overwhelm the potential improvements delivered by the quantum computer. Thus, using an appropriate embedding technique will be inevitable in future applications. Furthermore, as weakly correlated systems are more common in pharma, methods on a quantum computer must be compared to DFT in terms of accuracy and efficiency. If the method of choice is QPE, there is obvious advantage in obtaining the exact solution, while for ansatz-based approaches, a comparison to classical wavefunction based approaches might be appropriate. In terms of efficiency, quantum computers must not introduce a significant overhead compared to DFT so that the improvements in the accuracy of results will come at a reasonable cost. Despite the overall success of DFT, the constant call for better methods indicates that fault tolerant quantum computers have a significant contribution to make in many areas of chemistry.

It remains to note that even DFT is not widely applied throughout industrial computer-aided drug design workflows. We demonstrated the applicability of quantum computing algorithms for a realistic QM cluster approach, which similarly as the above described QM/MM method does indeed utilise quantum mechanics to gain insight into drug-protein binding mechanisms. However, both methods are usually used either for bespoke bits at the end of the computational drug design funnel or in academic pharmaceutical research. Current high throughput workflows are devised  to allow the processing of hundreds of thousands of structures with the limited classical computing resources available which renders even the usage of DFT with relatively cheap functionals unfeasible throughout most of the computational drug design pipeline. Rather than attempting to simply substitute existing steps in the workflows it will be a challenge for computational chemists and algorithmic researchers to re-think the computer-aided drug design processes while the hardware matures in the next years.

Yet, the thrilling perspectives for chemistry offered by quantum computing cannot be realised today. Even as different actors are racing to build and integrate larger and larger numbers of qubits \cite{wright_benchmarking_2019, Erhard2021, arute_quantum_2019, wu_strong_2021, collins2021ibm}, significant practical challenges in scaling-up the size of quantum computers remain. We have based our resource estimates on tomorrow's hardware that will have overcome these challenges, and today's algorithms. The high resource estimates thus show that tremendous effort must go also into improving algorithms and quantum error correction, improvements which have already allowed reduction in resources by orders of magnitude. As hardware developments and algorithmic requirements continue to draw closer to each other it is also important to not only improve resource estimates but look at implementing these aspects of the quantum computing stack in practice. One example of this is the recent demonstrations of quantum error correction on physical hardware \cite{chen_exponential_2021, nguyen_demonstration_2021, egan_fault-tolerant_2021, pino_demonstration_2021, postler2021demonstration, abobeih_fault-tolerant_2021}, but other levels must be developed as well. To unlock the potential of quantum computing, along with the physical engineering challenge one must address challenges across the entire stack.

\section*{Acknowledgements}

This work was performed as part of Astex's Sustaining Innovation Post-doctoral Program. We acknowledge funding from Innovate UK’s Sustainable Innovation Fund via SBRI. OC's contribution to the paper was in part supported by an Innovate UK grant (Quantum Enhanced Design for Materials and Chemistry, project number 105622). We thank David Plant and James R. Cruise for useful discussions and their valuable input to this paper.

\begin{appendices}

\section{Number of repetitions of QPE}\label{app:qperep}
As discussed in the main text, we will typically have to repeat the QPE procedure several times in order to obtain an estimate of the ground-state energy to the desired level of precision. In this appendix, we outline one possible method for deciding how many repetitions to perform and how to obtain an energy estimate from the results.

In order that our final energy estimate is to the desired accuracy with the desired probability, we repeat the phase estimation procedure $l$ times and take the median of the lowest $k$ measurement outcomes. These values are determined based on the details of the chemical system and calculation. Taking such a median reduces the effect of outliers.

In order to calculate $l$ and $k$, we make use of $P_{0}$, the probability that an eigenvalue estimate is within $2^{-t}$ of the true desired eigenvalue, assuming we are measuring the desired eigenstate to $t$ bits of precision. The derivation of this probability is given in Sec.~\ref{app:qpeprob}. We also make use of $\eta$, an estimate of the overlap probability of the initial state with the desired eigenstate. Finally, we require $P_{f}$, the probability of the error correction procedure failing on a single run of phase estimation, that is, the probability of an undetected error in magic state distillation or a logical error (see Sec.~\ref{sec:error-corrected resource estimation} for further details).

We then assume that all measurements of an \textit{excited} state where a logical error does not occur will result in an estimate \textit{above} the desired range, measurements of the \textit{ground} state where a logical error does not occur will result in an estimate \textit{within} the desired range with probability $P_{0}$, and all other measurements result in an estimate \textit{below} the desired range. The last choice in particular is likely to be worse than the true situation; this is deliberate so as to avoid underestimating the number of repetitions required. We then calculate the smallest value of $l$ and a corresponding value of $k$ such that the probability of the median estimate being within the desired range is at least the specified desired success probability, $P_{s}$. We find $k = 2\left \lfloor \frac{\eta m}{2} \right \rfloor + 1$ to perform well, though there can be improved values of $k$. We note that, choosing $k$ in this way, means the overall success probability is not necessarily larger if the true value of $\eta$ is greater than the value used to calculate $k$; as a result, it may be preferable to increase $l$ and/or $k$ to ensure the overall success probability is above the desired value for a full range of desired $\eta$ values.

\subsection{QPE probabilities}
\label{app:qpeprob}
We wish to find the probability of a single eigenvalue estimate being within $\epsilon = 2^{-t}$ of the true desired eigenvalue. We assume that each eigenvalue, $E_{j}$ satisfies $0 \leq E_{j} < 1$ and so can write the $j$th eigenvalue as
\begin{equation}
    E_{j} = \sum_{p=1}^{m} \phi_{jp}2^{-p} \delta_{j}2^{-m},
\end{equation}
where each $\phi_{jp}$ has a value of 0 or 1 and $0 \leq \delta_{j} < 1$. We will find it useful to define $b_{j}2^{-m} = \sum_{p=1}^{m} \phi_{jp} 2^{-p}$. We let $p_{l}$, where $-2^{m-1} < l \leq 2^{m-1}$, be the probability of obtaining measurements $(\theta_{l,m}, \theta_{l,m-1}, \ldots, \theta_{l,1})$ such that $\sum_{p=1}^{m}\theta_{lp}2^{m-p} = (b_{n}+l)\mod 2^{m}$, where $n$ is the index of the desired eigenvalue, assuming the desired eigenvalue is measured. Following standard manipulations~\cite{nielsen2002quantum,dobvsivcek2007arbitrary}, we see that
\begin{align}
    p_{l} &= \frac{\sin^{2}\{\pi[2^{m}E_{n}-(b_{n}+l)]\}}{\sin^{2}\{\pi[E_{n}-2^{-m}(b_{n}+l)]\}} \\
    &= \frac{1}{2^{2m}} \frac{\sin^{2}[\pi(\delta_{n}-l)]}{\sin^{2}[2^{-m}\pi(\delta_{n}-l)]} \\
    &= \frac{1}{2^{2m}}\frac{\sin^{2}[\pi(\delta_{n})]}{\sin^{2}[2^{-m}\pi(\delta_{n}-l)]} \\
     &= \frac{\sin^{2}[\pi(\delta_{n})]}{\pi^{2}(\delta_{n}-l)^{2}\mathrm{sinc}^{2}[2^{-m}\pi(\delta_{n}-l)]}.
\end{align}
This is a decreasing function of $m$. Taking the limit as $m \rightarrow \infty$, we see that
\begin{equation}
    p_{l} \geq \frac{\sin^{2}(\pi\delta_{n})}{\pi^{2}(\delta_{n}-l)^{2}}.
\end{equation}
We wish to to sum over these probabilities to find the probability, $P_{m-t}$, that the estimated eigenvalue is within a certain distance, $2^{-t}$, of the true eigenvalue, given we measure $m$ bits of precision. This is given by
\begin{align}
    P_{m-t} &= \sum_{l=-2^{m-t}+1}^{2^{m-t}} p_{l} \geq \vert c_{n} \vert^{2} \sum_{l=-2^{m-t}+1}^{2^{m-t}} \frac{\sin^{2}(\pi \delta_{n})}{\pi^{2}(\delta_{n}-l)^{2}} \\
    &= \vert c_{n} \vert^{2} \frac{\sin^{2}(\pi \delta_{n})}{\pi^{2}} \sum_{l=0}^{2^{m-t}-1} \left [ \frac{1}{(\delta_{n}+l)^{2}} + \frac{1}{(\delta_{n}-l-1)^{2}}\right ].
\end{align}
This is minimised when $\delta_{n} = \frac{1}{2}$. Therefore,
\begin{equation}
    P_{m-t} \geq \frac{8}{\pi^{2}} \sum_{l=0}^{2^{m-t}-1} \frac{1}{(2l+1)^{2}}.
\end{equation}
This has a number of terms that is exponential in the value of $m-t$; however, it will not be prohibitive to evaluate the sum for small values. For larger values, we can derive a bound which does not involve a sum. From the Basel problem, we have
\begin{equation}
    \sum_{l=0}^{\infty} \frac{1}{(2l+1)^{2}} = \frac{\pi^{2}}{8},
\end{equation}
and so
\begin{align}
    P_{m-t} &\geq \left ( 1 - \frac{8}{\pi^{2}}\sum_{l=2^{m-t}}^{\infty} \frac{1}{(2l+1)^{2}} \right ) \\
    &\geq \left ( 1 - \frac{8}{\pi^{2}}\int_{l=2^{m-t}-1}^{\infty} \frac{1}{(2l+1)^{2}} \mathrm{d}l \right ) \\
    &= \left ( 1 - \frac{4}{\pi^{2}} \frac{1}{(2^{m-t+1}-1)}\right ).
\end{align}




\end{appendices}


\providecommand{\latin}[1]{#1}
\makeatletter
\providecommand{\doi}
  {\begingroup\let\do\@makeother\dospecials
  \catcode`\{=1 \catcode`\}=2 \doi@aux}
\providecommand{\doi@aux}[1]{\endgroup\texttt{#1}}
\makeatother
\providecommand*\mcitethebibliography{\thebibliography}
\csname @ifundefined\endcsname{endmcitethebibliography}
  {\let\endmcitethebibliography\endthebibliography}{}

\end{document}